\documentclass[preprint,showpacs]{article}

\pdfoutput=1
\usepackage{graphicx,amsmath,amssymb,amsfonts}
\usepackage{enumerate,setspace}
\usepackage{psfrag}
\usepackage{xcolor}   
\usepackage[font={small}]{caption}
\usepackage{abstract}

\usepackage{makeidx}
\renewcommand{\a}{{\bf a}}
\usepackage{cite}
 \usepackage{hypernat}

\newcommand{\e}{{\rm e}}
\renewcommand{\O}{\mathcal{O}}

\newcommand{\ol}{\overline}

\newcommand{\tr}{E}

\renewcommand{\a}{{\bf a}}

\newcommand{\W}{{\cal W}}
\newcommand{\eq}{{\rm eq}}
\newcommand{\ad}{{\rm ad}}

\newcommand{\st}{{\rm stat}}

\newcommand{\cE}{\mathcal{E}}
\renewcommand{\v}{\upsilon}

\newcommand{\fig}[1]{\bf \large #1}
\usepackage{fullpage}

\newcommand{\ql}{ { \langle } }
\newcommand{\qr}{ { \rangle } }
\newcommand{\pl}{  \langle  }
\newcommand{\pr}{  \rangle }
\newcommand{\preq}{  \rangle_\eq }

\newcommand{\x}{ w(c)}
\newcommand{\xv}{ w(c,\v,r^2)}
\newcommand{\xmv}{ w(c,-\v,r^2)}
\newcommand{\tOne}{\tau_\eq(c)}
\newcommand{\tTwo}{\tau_{\rm sat} (v,r^2)}
\newcommand{\tOneA}{\tau_\eq}
\newcommand{\tTwoA}{\tau_{\rm sat}}
\newcommand{\sat}{{\rm sat}}

\newcommand{\micro}{\mbox{(microev. seascapes)}}
\newcommand{\macro}{\mbox{(macroev. seascapes)}}

	\usepackage{feynmp}
\newcommand{\EQ}{\begin{equation}}
\newcommand{\EE}{\end{equation}}
\newcommand{\EQA}{\begin{eqnarray}}
\newcommand{\EEA}{\end{eqnarray}}

\def\longrightharpoonup{\relbar\joinrel\rightharpoonup}
\def\longleftharpoondown{\leftharpoondown\joinrel\relbar}

\def\longrightleftharpoons{
  \mathop{
    \vcenter{
      \hbox{
	\ooalign{
	  \raise1pt\hbox{$\longrightharpoonup\joinrel$}\crcr
	  \lower1pt\hbox{$\longleftharpoondown\joinrel$}
	}
      }
    }
  }
}

\newcommand{\partialT}{\frac{\partial}{\partial t}}

\renewcommand{\d}{\text{d}}
\usepackage[font={small}]{caption}
\usepackage{rotating}

\usepackage{paralist}

\begin{document}

\begin{titlepage}
\title{Adaptive evolution of molecular phenotypes}
\author{Torsten Held$^{1}$, Armita Nourmohammad$^{1,2}$, Michael L\"assig$^{1}$}
\date{\small $^1$ Institut f\"ur Theoretische Physik, Universit\"at zu K\"oln, Z\"ulpicherstr. 77, \\
50937, K\"oln, Germany\\
$^2$ Joseph Henry Laboratories of Physics and Lewis-Sigler Institute for Integrative Genomics, \\ Princeton University, Princeton, NJ 08544, USA
}
\maketitle

\begin{abstract} 
Molecular phenotypes link genomic information with organismic functions, fitness, and evolution. Quantitative traits are complex phenotypes that depend on multiple genomic loci. In this paper, we  study the adaptive evolution of a quantitative trait under time-dependent selection,  which arises from environmental changes or through fitness interactions with other co-evolving phenotypes. We analyze a model  of trait evolution under mutations and genetic drift in a single-peak fitness seascape. The fitness peak performs a constrained random walk in the trait amplitude, which determines the time-dependent trait optimum in a given population. We derive analytical expressions for the distribution of the time-dependent trait divergence between populations and of the trait diversity within populations. Based on this solution, we develop a method to infer adaptive evolution of quantitative traits. Specifically, we show that the ratio of the average trait divergence and the diversity is a universal function of evolutionary time, which predicts the stabilizing strength and the driving rate of the fitness seascape. From an information-theoretic point of view, this function measures the macro-evolutionary entropy in a population ensemble, which determines the predictability of the evolutionary process. Our solution also quantifies two key characteristics of adapting populations: the cumulative fitness flux, which measures the total amount of adaptation, and the adaptive load, which is the fitness cost due to a population's lag behind the fitness peak. 
\end{abstract}
\thispagestyle{empty}
\end{titlepage}

\tableofcontents

\section{Introduction}

This is the second in a series of papers on the evolution of quantitative traits in biological systems~\cite{Nourmohammad2012}. We focus on molecular traits such as protein binding affinities or gene expression levels, which are mesoscopic phenotypes that bridge between genomic information and higher-level organismic traits. Such phenotypes are complex: they depend on tens to hundreds of constitutive genomic sites and are generically polymorphic in a population. Moreover, their evolution is often a strongly correlated process that involves linkage disequilibrium, i.e., allele associations due to incomplete recombination, and epistasis, i.e., fitness interactions, between constitutive sites. Hence, the evolutionary statistics of molecular quantitative traits have to go beyond traditional quantitative genetics~\cite{Fisher:1930wy,LANDE:1976tp,Barton:1989vm,Falconer:1989tg,Lynch:1998vx,Rice:1990vo,Hartl:1996td,Blumer:1972,Barton:1986,Wright:1935Jg}. Our aim is to derive {\em universal} phenotypic features of these processes, which decouple from details of a trait's genomic encoding and of the molecular evolutionary dynamics.

In this paper, we focus on the adaptive evolution of molecular traits, which involves mutations, genetic drift, and (partial) recombination of the trait loci. The adaptive dynamics take place on macro-evolutionary time scales and can generate large trait changes --- in contrast to micro-evolutionary processes based on standing trait variation in a population. Adaptive trait changes are driven by time-dependent selection on the trait values. Specifically, we consider the trait evolution in a single-peak {\em fitness seascape}~\cite{Mustonen:2009vu, KoppHermisson,deVladar:2011bs}, which has a moving peak described by a stochastic process in the trait coordinate. The time-dependence of the optimal trait value can have extrinsic or intrinsic causes; for example, the optimal expression level of a gene is affected by changes in the environment of an organism and by expression changes of other genes in the same gene network. These fitness seascape models have two fundamental parameters: the {\em stabilizing strength} and the {\em driving rate}, which measure the width  and the mean square displacement of the fitness peak per unit of evolutionary time. 

Here, we mainly focus on {\em macro-evolutionary} fitness seascapes, which have low driving rates compared to the diffusion of the trait by genetic drift~\cite{Mustonen:2009vu, Mustonen2008prl}. This kind of selection generates two complementary evolutionary forces. On short time scales, a single fitness peak acts as stabilizing selection, which constrains the trait diversity within a population as well as its divergence between populations. On longer time scales, the population trait mean follows the moving fitness peak, which generates an adaptive component of the trait divergence. In an  ensemble of populations with independent fitness peak displacements, these dynamics describe lineage-specific adaptive pressure. We discuss specific seascape models with continuous or punctuated adaptive pressure; that is, the fitness peak performs a constrained (Ornstein-Uhlenbeck) random walk or a Poisson jump process in the trait coordinate\footnote{The stochastic process of the fitness peak should not be confused with an Ornstein-Uhlenbeck dynamics of the trait mean that describes evolutionary equilibrium in a quadratic fitness landscape~\cite{Bedford:2009fy}.}. These stochastic processes define minimal non-equilibrium models for the adaptive evolution of a quantitative trait. In the limit case of a static fitness landscape, we recover the evolutionary equilibrium of quantitative traits under stabilizing selection, which has been the subject of a previous publication~\cite{Nourmohammad2012}. We also discuss the opposite regime of {\em micro-evolutionary} fitness seascapes, in which the evolution of the trait partially decouples from the movement of the fitness peak. 

Our model of adaptive trait evolution contains different sources of stochasticity: mutations establish trait differences between individuals within one population, reproductive fluctuations (genetic drift) and fitness seascape fluctuations generate trait differences between populations with time.  In macro-evolutionary fitness seascapes, these stochastic forces act on different time scales and define different statistical ensembles, similar to thermal and quenched fluctuations in the statistical thermodynamics of disordered systems. In section~\ref{sec:GenToTrait}, we derive stochastic evolution equations for the trait mean, the trait diversity, and the position of the fitness peak, which establish a joint dynamical  model for  the trait and the underlying fitness seascape over  macro-evolutionary time-scales. In section~3, we discuss the analytical solution of these models for a stationary ensemble of adapting populations. This ensemble has a time-independent trait diversity within populations, as well as a trait divergence between populations that depends on their divergence time. In section~4, we evaluate two important summary statistics of adaptive processes. The {\em genetic load}, which is defined as the difference between the maximum fitness and the mean population fitness, is shown to include a specific adaptive component, which results from the lag of the population behind the moving fitness peak. The {\em cumulative fitness flux} measures the amount of adaptation in a population over a macro-evolutionary period: it is zero at evolutionary equilibrium and  increases monotonically with the driving rate of selection~\cite{Mustonen2010}. Furthermore, we determine the {\em predictability} of trait values in one population given its distribution in another population, which is given by a suitably defined  entropy of the population ensemble under divergent evolution.

The statistical theory of this paper provides a new method to infer selection on a quantitative trait from diversity and time-resolved divergence data. Given these data in a family of evolving populations, we use the divergence-diversity ratio $\Omega (\tau)$ for different divergence times $\tau$ to determine stabilizing strength and driving rate of the underlying fitness seascape. These selection parameters, in turn, quantify the amount of conservation and adaptation in the evolution of the trait. Unlike previously used measures of trait evolution~\cite{Gilad:2006kt, Bedford:2009fy}, the divergence-diversity ratio is universal: it is determined by stabilizing strength and driving rate of the fitness seascape and on the evolutionary distance between populations, but it depends only weakly on the trait's constitutive sites, on the amount of recombination between these sites, and on details of the fitness dynamics. In contrast to most sequence evolution tests, the $\Omega$ test does not require the gauge of a neutrally evolving ``null trait''. We discuss this test statistics in section~\ref{sec:Inference}. For the reader not interested in technical details, Figure~\ref{fig:summary} 
provides a fast track through the preceding sections.

The statistical theory of this paper provides a new method to infer selection on a quantitative trait from diversity and time-resolved divergence data. Given these data in a family of evolving populations, we use the divergence-diversity ratio $\Omega (\tau)$ for different divergence times $\tau$ to determine the stabilizing strength and the driving rate of the underlying fitness seascape. These selection parameters, in turn, quantify the amount of conservation and adaptation in the evolution of the trait. Unlike previously used measures of trait evolution~\cite{Gilad:2006kt, Bedford:2009fy}, the divergence-diversity ratio is universal: it depends on the stabilizing strength and the driving rate of the fitness seascape as well as on the evolutionary distance between populations, but it is largely independent of the trait's constitutive sites, of the amount of recombination between these sites, and of the details of the fitness dynamics. In contrast to most sequence evolution tests, the $\Omega$ test does not require the gauge of a neutrally evolving ``null trait''. We discuss this test statistics in section~\ref{sec:Inference}. For the reader not interested in technical details, Figure~\ref{fig:summary} 
provides a fast track through the preceding sections.

\section{Evolutionary dynamics of quantitative traits}
\label{sec:GenToTrait}

In this section, we develop minimal models for the adaptive evolution of quantitative traits in a fitness seascape. We first review the diffusion dynamics for the trait mean and the diversity under genetic drift and mutations in a given fitness landscape, which have been derived in a previous paper~\cite{Nourmohammad2012}. Second, we introduce simple stochastic models for the dynamics of selection, which promote fitness landscapes to fitness seascapes. We then combine the dynamics of trait and selection to  a joint, non-equilibrium evolutionary model.

\subsection{Diffusion equations for trait mean and diversity}

Our model for quantitative traits is based on a simple additive map from genotypes to phenotypes. The trait value $E$ of an individual depends on its genotype $\a=(a_1, \dots,a_\ell)$ at~$\ell$ constitutive genomic sites, 
\EQ
\tr(\mathbf a) = \sum_{i=1}^\ell E_i \sigma_i, \hspace{1cm}\text{with } \sigma_i \equiv 
\begin{cases} \begin{array}{ll}  1, & \text{if }a_i=a_i^*,\\ 0, & \text{otherwise.} \end{array}
\end{cases}
\label{map}
\EE
Here, the trait is measured from its minimum value, and $E_i>0$ is the contribution of a given site $i$ to the trait value. We assume a two-allele genomic alphabet and $a_i^*$ denotes the allele conferring the larger phenotype  at site $i$. The extension to a four-allele alphabet is straightforward.  The genotype-phenotype map  (\ref{map}) defines the allelic trait average $\Gamma_0$ and the trait span $E_0^2$,
\EQ
\Gamma_0=\frac{1}{2}\sum_{i=1}^\ell E_i, \hspace{1cm}E_0^2=\frac{1}{4}\sum_{i=1}^\ell E_i^2.
\label{E0}
\EE
Quantitative traits have a sufficient number of constitutive loci to be generically polymorphic in a population, although most individual genomic sites are monomorphic. The distribution of trait values in a given population, $\W(E)$, is often approximately Gaussian \cite{deVladar:2011bs,LANDE:1976tp,Nourmohammad2012}.  Hence, it is well characterized by its mean and variance, 
\EQA
\Gamma \equiv \ol E & = & \int \! \d E \, E \, \W (E)  , 
\nonumber \\
\Delta \equiv \ol{ (E-\Gamma)^2} & = & \int \! \d E \, (E - \Gamma)^2 \, \W(E) , 
\label{eq:Stat}
\EEA
where overbars denote averages over the trait distribution $\W (E)$ within a population. The variance $\Delta$ will be called the trait diversity; in the language of quantitative genetics, this quantity equals the total heritable variance including epistatic effects. 

We consider the evolution of the trait $E$ under genetic drift, genomic mutations, and natural selection, which is given by a trait-dependent fitness seascape $f(E,t)$ that changes on macro-evolutionary time scales. This process is illustrated in Fig.~1: At a given evolutionary time, the trait distribution in a population has mean $\Gamma (t)$, diversity $\Delta (t)$, and is positioned at a distance $\Lambda (t) \equiv \Gamma (t) - E^*(t)$ from the optimal trait value. The trait distribution follows the moving fitness peak, building up a trait divergence 
\EQ
D^{(1)}(t, \tau) = (\Gamma (t) - \Gamma (t - \tau))^2 
\label{def:D}
\EE
between an ancestral population at time $t - \tau$ and its descendent population at time $t$ in a given lineage. In the same way, we can define the trait divergence between two descendent populations at time $t$ that have evolved independently from a common ancestor population at time $t - \tau/2$, 
\EQ
D^{(2)}(t,\tau)= (\Gamma_1(t)-\Gamma_2(t) )^2.
\label{D2}
\EE
In a suitably defined ensemble of parallel-evolving populations, the expectation values of these divergences, $\langle D^{(\kappa)}(\tau) \rangle$ ($\kappa = 1,2$), depend only on the divergence time $\tau$. The asymptotic divergence for long times is just twice the trait variance across populations, 
\EQ
\lim_{\tau \to \infty} \langle D^{(\kappa)}(\tau)\rangle  = 2\langle (\Gamma -  \langle \Gamma \rangle)^2 \rangle 
\hspace{1cm}(\kappa = 1,2). 
\EE
In particular, the quantity $E_0^2$ defined in (\ref{E0}) is the trait variance in an ensemble of random genotypes, which results from neutral evolution (with averages $\langle \dots \rangle_0$ marked by a subscript) at low mutation rates, 
$E_0^2 = \lim_{\mu \to 0} \langle (\Gamma -  \langle \Gamma \rangle_0)^2 \rangle_0$. For finite times, however, the statistics of the single-lineage divergence $D^{(1)}$ and the cross-lineage divergence $D^{(2)}$ differ from each other in an adaptive process. As we will discuss in detail below, this is a manifestation of the non-equilibrium evolutionary dynamics in a fitness seascape. In contrast,  evolutionary equilibrium in a fitness landscape dictates $\langle D^{(1)}(\tau) \rangle_\eq=\ql D^{(2)}(\tau)\qr_\eq$ by detailed balance.

\begin{figure}
\centering
\includegraphics[width=.95\textwidth]{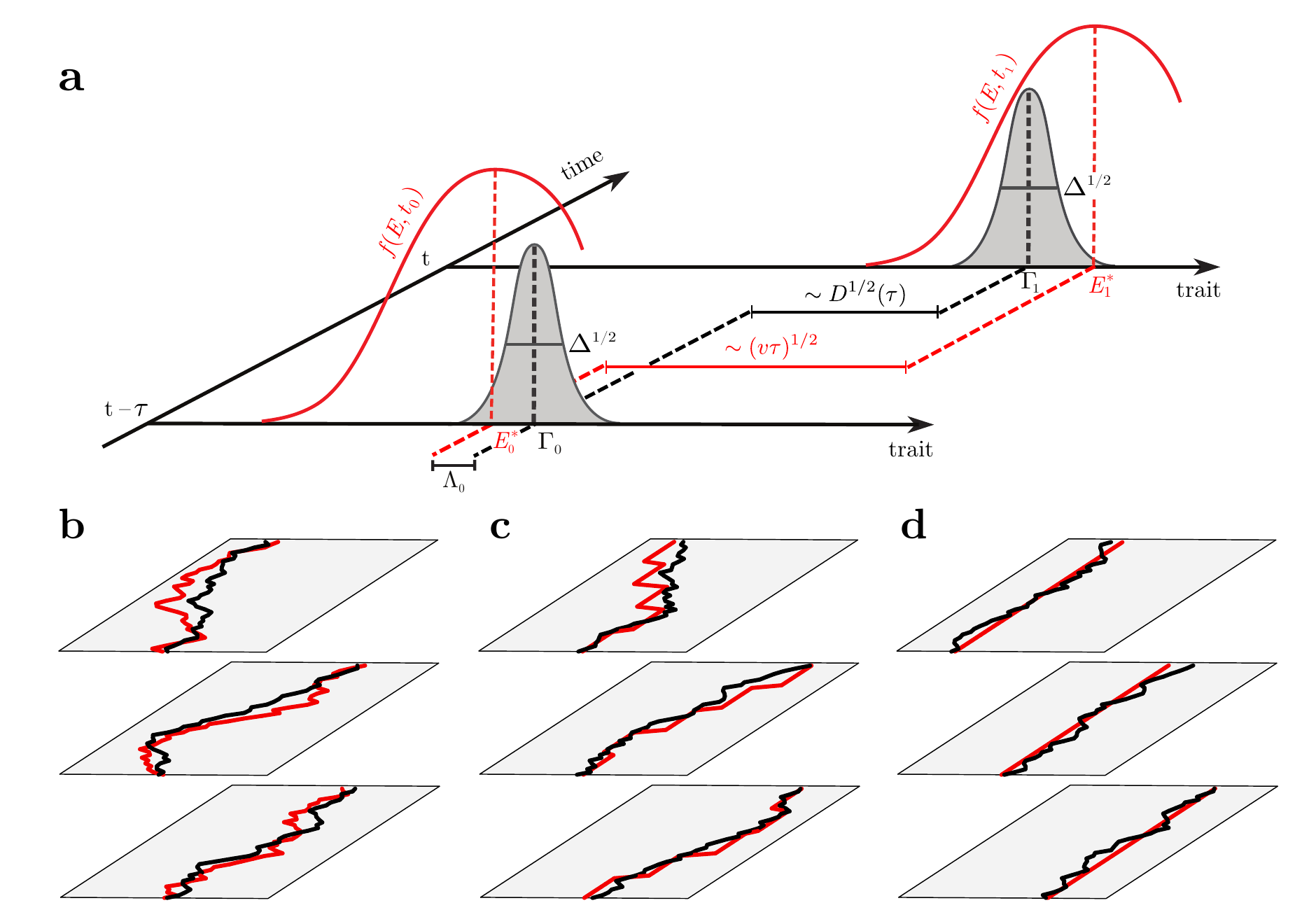}

\caption{{\bf Adaptive evolution of a quantitative trait.} (a) Evolution of the distribution of trait values $\W(E,t)$ (gray curves) in a given population subject to a single-peak fitness seascape $f(E,t)$ (red curves). At a given  time $t$, the population has a trait  distribution $\W(E,t)$ with mean $\Gamma (t)$ and diversity $\Delta (t)$, and is positioned at a distance $\Lambda (t) = \Gamma (t) - E^*(t)$ from the fitness peak. The population follows the moving fitness peak and builds up a trait divergence $D^{(1)}(t,\tau) = (\Gamma(t) - \Gamma (t - \tau))^2$ between the ancestral state at time $t -\tau$ and the descendent state at time $t$. The divergence $D^{(2)}(t,\tau)$ between two descendent populations with a common ancestor at time $t - \tau/2$ can be defined in an analogous way; see eqs.~(\ref{def:D}) and (\ref{D2}). 
(b--d) Evolutionary population ensembles, each represented by three sample populations. In a given population, a realization of a single-peak fitness seascape specifies a lineage-specific optimal trait value that depends on evolutionary time, $E^* (t)$ (red line). The population mean trait, $\Gamma (t)$ (black line), adapts to the moving fitness peak with additional lineage-specific fluctuations due to mutations and genetic drift. 
The adaptive process is shown for three cases of fitness seascapes: 
(b)~Diffusive fitness seascape: Incremental changes in the optimal trait value reflect adaptive pressure caused by continuous ecological changes. The function $E^* (t)$ follows an Ornstein-Uhlenback random walk in the trait coordinate. 
(c{)}~Punctuated fitness seascape: Sudden changes in the optimal trait value reflect adaptive pressure caused by major, discrete ecological events. The function $E^*(t)$ is described by a Poisson jump process in the trait coordinate. We show that both types of fitness seascapes lead to a solvable, non-equilibrium joint statistics of $\Gamma$ and $E^*$. 
(d)~Fitness landscape: Each population has a time-independent optimal trait value $E^*$ and reaches an evolutionary (selection-mutation-drift) equilibrium~\cite{Nourmohammad2012}. 
}
\label{fig:summary}
\end{figure}

As shown in a previous companion paper~\cite{Nourmohammad2012}, the evolutionary dynamics of a quantitative trait in a fitness seascape can be described in good approximation by diffusion equations for its mean and its diversity, 
\EQA
\frac{\partial}{\partial t} Q(\Gamma,t \, |\, F_1) & = & 
\left[\frac{g^{\Gamma \Gamma}}{2N}\frac{\partial^2}{\partial\Gamma^2} 
- \frac{\partial}{\partial\Gamma} \left( m^\Gamma + g^{\Gamma \Gamma}  \frac{\partial F_1(\Gamma,t)}{\partial \Gamma} \right) \right ] Q(\Gamma,t \, | \,  F_1),
\label{diffGamma}
\\ \nonumber \\
\frac{\partial}{\partial t} Q(\Delta,t \, | \,  F_2) & = & 
\left[\frac{g^{\Delta \Delta}}{2N}\frac{\partial^2}{\partial\Delta^2} 
- \frac{\partial}{\partial\Delta} \left( m^\Delta + g^{\Delta \Delta}  \frac{\partial F_2(\Delta,t)}{\partial \Delta} \right) \right ] Q(\Delta ,t \, | \,  F_2),
\label{diffDelta}
\EEA
which are projections of the Kimura diffusion equation \cite{Kimura:1964,Ewens} from genotypes onto the phenotype space. The distributions $ Q(\Gamma,t \, |\, F_1)$ and $Q(\Delta,t \, | \,  F_2)$ describe an ensemble of populations evolving in the same fitness seascape $f(E,t)$. These dynamics involve the fitness seascape components 
\EQA
F_1 (\Gamma, t) & = & f (\Gamma, t) + f'' (\Gamma, t) \times \int \! \d \Delta \,\Delta \, Q_2 (\Delta, t \, | \, F_2) ,
\label{FGamma}\\
F_2 (\Delta, t) & = & \Delta \times \int \! \d \Gamma \, f'' (\Gamma, t) \, Q(\Gamma, t \, | \, F_1) ,
\label{FDelta}
\EEA 
which are projections of the mean population fitness 
\EQ
\ol f (t) \equiv \int  \d E \, f(E,t) \, \W(E,t)  = f(\Gamma,t)+\frac{1}{2}\Delta f^{\prime\prime}(\Gamma,t) + \dots
\EE 
onto the marginal variables $\Gamma$ and $\Delta$. Genetic drift enters through the diffusion coefficients $g^{\Gamma\Gamma} = \langle \Delta \rangle \equiv \int \! \d \Delta \, \Delta \, Q (\Delta, t \, | \, F_2)$ and $g^{\Delta \Delta} = 2\Delta^2$, mutations through the mutation coefficients $m^\Gamma = -2\mu (\Gamma-\Gamma_0)$ and $m^\Delta = -4\mu(\Delta-E_0^2)- \Delta/N$; these coefficients depend on the effective population size $N$ and the point mutation rate $\mu$. The diffusion equations (\ref{diffGamma}) and (\ref{diffDelta}) are coupled through the fitness components (\ref{FGamma}) and (\ref{FDelta}) and through the diffusion coefficient $g^{\Gamma\Gamma}$. In the absence of direct selection on the trait mean, i.e.~for  $F_1(\Gamma,t)=0$, eq.~(\ref{diffGamma}) describes a ``quasi-neutral'' diffusion of the trait mean, which depends the full drift term  $g^{\Gamma \Gamma}=\langle \Delta \rangle(c)$ under selection. This dynamics defines a characteristic time scale 
\EQ
\tilde \tau \equiv \frac{2N E_0^2}{\langle \Delta \rangle }.
\label{tautilde}
\EE
Recombination between the trait loci induces a crossover between selection on entire genotypes and selection on individual alleles~\cite{Neher:2009hva,shraiman:2012}. This affects the form of the diffusive dynamics of $\Delta$ and, hence, the coefficients of the diffusion equation for $\Gamma$. However, these changes have only a small effect on the statistics of $\Gamma$ over a wide range of evolutionary parameters~\cite{Nourmohammad2012}. 

In the special case of a time-independent fitness landscape $f(E)$, the diffusive dynamics of trait mean and diversity leads to  evolutionary equilibria of a Boltzmann form~\cite{Nourmohammad2012}, 
\EQA 
Q_\eq (\Gamma \, | \, F_1) & = & \frac{1}{Z_\Gamma}\,  \tilde Q_0 (\Gamma) \, \exp[2N F_1 (\Gamma)], 
\label{QeqGamma}
\\
Q_\eq (\Delta \, | \, F_2) & = & \frac{1}{Z_\Delta}\,  Q_0 (\Delta) \, \exp[2N F_2 (\Delta)], 
\label{QeqDelta}
\EEA
where $Z_\Gamma$ and $Z_\Delta$ are normalization constants. The equilibrium distributions under selection build on the quasi-neutral distribution of the trait mean, $\tilde Q_0 (\Gamma) \sim \exp[ - 2 \mu N (\Gamma - \Gamma_0)^2 / \langle \Delta \rangle]$, and on the neutral diversity distribution $Q_0 (\Delta)$ (see also section \ref{chap:delta}). We note that evolutionary equilibrium in a static fitness landscape is limited to the marginal distributions $Q_\eq (\Gamma \, | \, F_1)$ and $Q_\eq (\Delta \, | \, F_2)$, while the joint distribution $Q (\Gamma,\Delta|f)$ reaches a non-equilibrium stationary state~\cite{Nourmohammad2012}. In the limit of low mutation rates, the Boltzmann distribution (\ref{QeqGamma}) describes an asymptotic selection-drift equilibrium $Q_\eq (E | F_1) \sim Q_0 (E) \exp[2N F(E)]$; the trait values $E$ are predominantly monomorphic in a population and they change by substitutions at individual trait loci~\cite{Berg:2004dz, Lassig:2007iq,Nourmohammad2012}. 

\subsection{Stochastic seascape models} 
\label{sec:seascape}
For a generic fitness seascape $f(E,t)$, the diffusion equations (\ref{diffGamma}) and (\ref{diffDelta}) do not have a closed analytical solution. At the same time, we are often not interested in the detailed history of fitness peak displacements and the resulting trait changes. To describe generic features of adaptive processes, we now introduce solvable stochastic models of the seascape dynamics and link broad features of these models to statistical observables of adapting populations. 

In this paper, we restrict our analysis to single-peak fitness seascapes of the form
\EQ
f(E,t) = f^* - c_0 \big ( E - E^* (t) \big )^2.  
\label{fE}
\EE
Despite its simple form, the fitness function (\ref{fE})  covers a broad spectrum of interesting selection scenarios~\cite{Nourmohammad2013}. For constant trait optimum $E^*$, it is a time-honored model of {\em stabilizing selection}.
\cite{Fisher:1930wy,Lynch:1998vx,Barton:1986,Nourmohammad2013,Berg:2004dz,ML05,Barton:2009di,Barton:2009genetics,Neher:2011wc,deVladar:2011bs}. Nearly all known examples of empirical fitness landscapes for molecular quantitative traits are of  single-peak~\cite{Poelwijk:2011ce}  or mesa-shaped~\cite{Gerland03,ML05, Kinney:2008tb,Wylie:2011io,Hermsen:2012fz} forms. Mesa landscapes describe directional selection with diminishing return: they contain a fitness flank on one side of a characteristic ``rim'' value $E^*$ and flatten to a plateau of maximal fitness on the other side. Furthermore, trait values on the fitness plateau tend to be encoded by far fewer genotypes than low-fitness values. This differential coverage of the genotype-phenotype map turns out to generate an effective second flank of the fitness landscape, which makes our subsequent theory applicable to mesa landscapes as well~\cite{Nourmohammad2013}. We refer to the scaled parameter
\EQ
c = 2 N E_0^2 c_0
\label{c}
\EE
as the {\em stabilizing strength} of a fitness landscape. This dimensionless quantity has a simple interpretation: it equals the ratio of the neutral trait variance $E_0^2$ and the weakly deleterious trait variance around the fitness peak, which, by definition, produces a fitness drop $\leq 1/(2N)$ below the maximum $f^*$. As shown in ref.~\cite{Nourmohammad2012}, the mutation-selection-drift dynamics of a quantitative trait in a single-peak fitness landscape leads to evolutionary equilibrium with a characteristic equilibration time 
\EQA
\tau_\eq (c) & = & \frac{1}{\mu + c \tilde \tau^{-1} (c)} 
\simeq \left \{ \begin{array}{ll} 
\mu^{-1} & \mbox{ for $c \lesssim 1$,}
\\
(c \tilde \tau (c))^{-1} & \mbox{ for $c \gtrsim 1$,}
\end{array} \right. 
\label{taueq}
\EEA
where $\tilde \tau(c)$ is the quasi-neutral drift time defined in eq.~(\ref{tautilde}). 

For time-dependent $E^*(t)$, eq.~(\ref{fE}) becomes a fitness seascape model \cite{Mustonen:2009vu, KoppHermisson,deVladar:2011bs}. At any given evolutionary time, this model describes stabilizing selection of strength $c$ towards an optimal trait value~$E^* (t)$. In addition, the changes of~$E^*(t)$ over macro-evolutionary periods introduce directional selection on the trait and generate adaptive evolution. The form (\ref{fE}) of a fitness seascape assumes the stabilizing strength $c$ to remain constant over time. As discussed in section~\ref{2.3}, this assumption leads to an important computational simplification: only the trait mean $\Gamma$ adapts to the moving fitness peak, while the diversity $\Delta$ remains at evolutionary equilibrium. However, generalizing of our model to a time-dependent stabilizing strength $c(t)$ is straightforward and is briefly discussed below. We consider two minimal models of seascape dynamics:

\paragraph{Diffusive fitness seascapes.} In this model, the fitness optimum $E^*(t)$ performs an Orn\-stein-Uhlenbeck random walk with diffusion constant $\v_0$, average value $\cE$ and stationary mean square deviation $r_0^2$.  The scaled parameters
\EQ
\v = \frac{\v_0}{E_0^2}, 
\hspace{1cm}
r^2 = \frac{r_0^2}{E_0^2},
\label{v}
\EE 
will be called the {\em driving rate} and the {\em driving span} of a fitness seascape. Different realizations of this random walk with the same set of parameters are shown in Fig.~\ref{fig:summary}(b). The distribution of optimum trait values, $R(E^*, t)$, follows a diffusion equation, 
\EQ
\frac{\partial}{\partial t} R(E^*,t) =  {\v E_0^2} \, \frac{\partial}{\partial E^*} \left [\frac{\partial}{\partial {E^*}} +  \frac{1}{r^2 E_0^2} (E^* - \cE) \right ] R(E^*,t).
\label{eq:FP_R_general}
\EE
This dynamics leads to a seascape ensemble, which is characterized by an expected peak divergence  
\EQ
\big \langle (E^*(t) - E^* (t + \tau))^2 \big \rangle =2 r^2 E_0^2 \left (1 - {\rm e}^{- \tau / \tau_{\rm sat} (\v, r^2)} \right )
\label{dEstar}
\EE
with the saturation time 
\EQ
\tau_{\rm sat}(\v, r^2)  = \frac{r^2}{ \v},
\label{tausat}
\EE
and by an equilibrium distribution
\EQ
R_\eq (E^*) = \frac{1}{\sqrt{2 \pi r^2 E_0^2} } \, \exp \left[ -\frac{1}{2} \frac{(E^* - \cE)^2}{ r^2 E_0^2} \right ]
\label{Req}
\EE
of optimal trait values. Diffusive seascapes models of this form describe continuous adaptive pressure due to incremental ecological changes that affect the optimal trait value $E^* (t)$. We assume that typical optimal trait values fall into the neutral trait repertoire given by eq.~(\ref{E0}), which implies that the scaled driving span $r^2$ is at most of order 1.

\paragraph{Punctuated fitness seascapes.} In this model, the fitness optimum performs a Poisson jump process with jump rate $\tau^{-1}_{\rm sat}(v,r^2)= \v /r^2$, by which successive values of $E^*$ are drawn independently from the distribution~$R_\eq (E^*)$, given by~(\ref{Req}). Different realizations of this process are shown in Fig.~\ref{fig:summary}(c). Fitness jumps may result from discrete ecological events such as major migrations or speciations. The Poisson jump process is described by the evolution equation 
\EQ
\frac{\partial}{\partial t} R(E^*,t) =  \frac{\v}{r^2} \, \big [ R_\eq (E^*) - R(E^*, t)]. 
\label{jump}
\EE
It has the same time-dependent expected peak divergence~(\ref{dEstar}) and the same equilibrium distribution~(\ref{Req}) as the diffusion process~(\ref{eq:FP_R_general}) with  same driving parameters~(\ref{v}). The difference between the jump process and the diffusion process lies in the anomalous scaling of higher moments, 
\EQA
\big \langle (E^*(t) - E^* (t + \tau))^k \big \rangle & \sim & E_0^k r^{k-2} \v \tau 
\hspace{1cm} \mbox{ for $k = 2, 4, \dots$  and $\tau \ll \tau_{\rm sat} (\v, r^2)$.} 
\label{higher_moments}
\EEA
This scaling is shared by simple models of turbulence; see, e.g., ref.~\cite{Lassig:turb}. 

\bigskip
In both types of fitness seascape, we distinguish two dynamical selection regimes: 

\begin{itemize}
\item {\em Macro-evolutionary} fitness seascapes are defined by the condition $\tTwo \gtrsim \tOne $. As discussed in detail below, such seascapes keep the trait mean always close to equilibrium and induce an adaptive response linear in the driving rate $\v$.  The limit $\v \to 0$ describes an ensemble of {\em quenched} population-specific fitness {\em landscapes} with a distribution of optimal trait values given by eq.~(\ref{Req}); see Fig.~1(d). 

\item {\em Micro-evolutionary} fitness seascapes have $\tTwo \lesssim \tOne $ and delineate a regime of reduced adaptive response, where the evolution of the trait mean gradually decouples from that of the fitness seascape. In the asymptotic fast-driving regime  ($\v \gg r^2 /\tau_\eq(c)$), the adaptation of the trait is completely suppressed. In this regime, we can average over the fitness fluctuations and describe the macro-evolution of the trait in terms of an effective fitness landscape with an optimal trait value $\mathcal E$. 
\end{itemize}

\subsection{Joint dynamics of trait and selection}
\label{2.3}

We now combine the diffusive dynamics of quantitative traits in a given fitness seascape, which is given by eqs.~(\ref{diffGamma}) and (\ref{diffDelta}), and the seascape dynamics (\ref{eq:FP_R_general}) or  (\ref{jump}) into a stochastic model of adaptive evolution. The statistical ensemble generated by this model is illustrated in Fig.~\ref{fig:summary}(b-d): Each population evolves in a specific realization of the fitness seascape, which is given by a history of peak values $E^*(t)$. Its trait mean~$\Gamma (t)$ follows the moving fitness peak with fluctuations due to mutations and genetic drift. The ensemble of populations contains, in addition, the stochastic differences between realizations of the fitness seascape. The statistics of this ensemble involves combined averages over both kinds of fluctuations, which are denoted by angular brackets $\ql ... \qr$.

The population ensemble can be described by a joint distribution of mean and optimum trait values, $Q(\Gamma, E^*, t) = Q(\Gamma, t \, | \, E^*) R(E^*,t)$. Using eqs.~(\ref{diffGamma}) and  (\ref{eq:FP_R_general}) together with the projection  of the fitness seascape, 
\EQ
F_1 (\Gamma \, | \, E^*) = f^* - \frac{c}{N E_0^2} \ql \Delta \qr - \frac{c}{E_0^2} \big (\Gamma - E^* \big )^2,
\label{F1}
\EE
given by  eqs.~(\ref{FGamma}) and~(\ref{fE}), we obtain the evolution equation for the joint distribution in a diffusive seascape, 
\EQA
{\frac{\partial}{\partial t} Q(\Gamma, E^*,t)   } & = & 
\left [\frac{g^{\Gamma \Gamma}}{2N}\frac{\partial^2}{\partial\Gamma^2} 
- \frac{\partial}{\partial\Gamma} 
\left( m^\Gamma  - g^{\Gamma \Gamma} \frac{2 c}{E_0^2}  (\Gamma - E^* ) \right ) \right .
 + \left . \frac{\v}{  E_0^2} \frac{\partial^2}{\partial E^{* 2}}  +  \frac{\v}{r^2} \frac{\partial}{\partial {E^*}}(E^* - \cE)
\right ]  Q(\Gamma, E^*,t),
\nonumber \\ 
\label{diffGammaEstar} 
\EEA
with $g^{\Gamma\Gamma} = \langle \Delta \rangle$ and $m^\Gamma = -2\mu (\Gamma-\Gamma_0)$. Note that the differential operator in eq.~(\ref{diffGammaEstar}) is asymmetric: the trait optimum $E^*$ follows an independent stochastic dynamics, but the trait mean $\Gamma$ is coupled to $E^*$. This asymmetry reflects the causal relation between selection and adaptive response: the trait mean $\Gamma (t)$ follows the moving fitness peak, as shown in Fig.~\ref{fig:summary}(b,c). As a consequence, the joint evolution equation~(\ref{diffGammaEstar}) leads to a non-equilibrium stationary distribution $Q_{\rm stat} (\Gamma, E^*)$, although the marginal seascape dynamics~(\ref{eq:FP_R_general}) reaches an equilibrium state. Only in the fitness landscape limit ($\v \to 0$), the evolution of the trait mean reaches evolutionary equilibrium. In the next section, we will obtain explicit solutions for the non-equilibrium distribution~$Q_\st (\Gamma, E^*)$ and its equilibrium limit. The case of a punctuated fitness seascape is treated in Appendix~\ref{chap:solveG}, where we solve the Langevin equations for~$\Gamma$ and~$E^*$ to obtain the first and second moments of~$Q(\Gamma, E^*,t)$. 

The trait diversity evolves under the projected fitness function
\EQ
F_2 (\Delta \, | \, c)  = - \frac{c}{N E_0^2} \, \Delta,
\label{F2}
\EE
given by eqs.~(\ref{fE}) and~(\ref{FDelta}). In a fitness seascapes with a constant stabilizing strength $c$, this function is time-independent.  The dynamics  of the trait diversity~(\ref{diffDelta})  decouples from the adaptive evolution of the trait mean and leads an evolutionary equilibrium~$Q_\eq (\Delta \, | \, c)$ of the form~(\ref{QeqDelta}). As detailed in Section 3.3, the equilibrium assumption for  the trait diversity holds for most adaptive processes in a fitness seascape of the form (\ref{fE}). However, we can generalize our seascape models to include a time-dependent stabilizing strength $c(t)$. This leads to generic adaptive evolution of both, $\Gamma$ and $\Delta$, which is described by a coupled non-equilibrium stationary  distribution   
$Q_{\rm stat} (\Gamma, \Delta,E^*, c)$.

\section{Adaptive evolution in a single-peak fitness seascape}
In this section, we develop the key analytical results of this paper. We provide an explicit solution for the non-equilibrium joint distribution of mean and optimal trait in a diffusive  seascape, $Q_\st (\Gamma, E^*)$; the case of punctuated seascapes is treated in Appendix~\ref{chap:solveG}. These solutions describe a stationary ensembles of adapting populations. We derive an expression for the expected time-dependent trait divergence in these ensembles, which holds for both seascape models. Finally, we juxtapose the adaptive behavior of the trait mean with the equilibrium statistics of the trait diversity, which emerges in good approximation for most fitness seascape of constant stabilizing strength. 
Our analytical results are supported by simulations for diffusive and punctuated fitness seascapes.

\subsection{Stationary distribution of mean and optimal trait}
\label{chap:gamma}

In a diffusive fitness seascape, the evolution equation (\ref{diffGammaEstar}) has a stationary solution of bivariate Gaussian form, 
\EQA
Q_\st (\Gamma,E^*) = 
\frac{1}{Z} \exp\left[ -\frac{1}{2} \begin{pmatrix} \hat \Gamma  \\ \hat E^* \end{pmatrix}^T \mathbf \Sigma^{-1} \begin{pmatrix} \hat \Gamma  \\ \hat E^* \end{pmatrix}  \right],
\label{eq:ansatz1}
\EEA
with the expectation values
\EQA
\left ( 
\begin{array}{c} \langle \Gamma \rangle \\ \langle E^* \rangle 
\end{array} \right )  
& \equiv &  
\int \! \d \Gamma \d E^*  \,\begin{pmatrix}  \Gamma  \\  E^* \end{pmatrix} 
  Q_\st (\Gamma,E^*)
\nonumber \\
& = & 
\left ( 
\begin{array}{c} \x\, \mathcal E + (1-\x) \, \Gamma_0 \\ 
\mathcal E 
\end{array} \right ) ,
\label{ave}
\EEA
and the covariance matrix 
\EQA
\mathbf \Sigma  = \begin{pmatrix}  \ql \hat \Gamma^2 \qr  & { \ql }\hat \Gamma \hat E^*{ \qr }  \\ { \ql }\hat \Gamma \hat E^*{ \qr }  & \langle \hat E^{*2} \rangle \end{pmatrix} 
& \equiv & 
\int \! \d \Gamma \d E^*  \,\begin{pmatrix}  \hat \Gamma^2   & \hat \Gamma \hat E^*  \\ \hat \Gamma \hat E^* & \hat E^{*2} \end{pmatrix} 
  Q_\st (\Gamma,E^*)
\nonumber \\
& = & E_0^2 \begin{pmatrix}  (1/2c)\, \x +r^2 \x \xv& r^2 \xv  \\ r^2\xv& r^2 \end{pmatrix} ,
\label{eq:Sigma.full}
\EEA 
where $\hat \Gamma \equiv \Gamma - \langle \Gamma \rangle$ and $\hat E^* \equiv E^* - \mathcal E$. The stationary distribution  $Q_\st (\Gamma,E^*)$ depends on the  parameters that characterize the fitness seascape: the stabilizing strength $c$, the driving rate $\v$, and the relative driving span $r^2$, which are defined in eqs.~(\ref{c}) and (\ref{v}). Together with the effective population size $N$ and the point mutation rate $\mu$, these parameters determine the characteristic time scales of evolution in a fitness seascape, the equilibration time $\tau_\eq(c)$ and the saturation time of fitness fluctuations, $\tau_\sat (\v,r^2)$; see eqs.~(\ref{taueq}) and (\ref{tausat}). 
The function 
\EQA
\xv &\equiv& \frac{c\ql \delta \qr}{c\ql \delta \qr+2\theta+N\v/r^2} 
= \frac{\tau_\eq^{-1}(c) - \mu}{\tau_\eq^{-1}(c) - \mu + \tau_{\rm sat}^{-1} (\v, r^2)},
\label{xv} 
\EEA
and its equilibrium limit $w (c) \equiv w(c, \v \! = \! 0, r^2)$ govern the coupling between the mean and optimal trait. The mutation rate $\mu$ is the inverse of the neutral timescale $\tau_\eq(0)=\mu^{-1}$. These functions depend on the scaled diversity $\langle \delta \rangle \equiv \langle \Delta \rangle / E_0^2$,  which is given  in ref.~\cite{Nourmohammad2012}, eqs.~(68)~--~(73), and is restated below in eq.~(\ref{deltaconstraint}). For traits under substantial selection ($c \gtrsim 1$), we can distinguish two dynamical regimes: In macro-evolutionary fitness seascapes, where $\tau_{\rm sat} (\v, r^2) \gtrsim \tau_\eq (c) \approx 2N / (\langle \delta \rangle c)$, this coupling remains close to the equilibrium value $w(c) \approx 1$;  micro-evolutionary fitness fluctuations, which have $\tau_{\rm sat} (\v, r^2) \lesssim \tau_\eq (c)$, induce a partial decoupling of mean and optimal trait. 

We can also express this crossover in terms of  the  average square distance between trait mean in the population and optimal trait  of the underlying fitness seascape,
\EQA
\langle \Lambda^2 \rangle & \equiv &  
\int \! \d \Gamma \d E^* \, (\Gamma - E^*)^2 
\, Q_\st (\Gamma,E^*). \label{eq:defLambda}
\EEA
The analytical solution for the scaled quantity $\langle \lambda^2 \rangle \equiv \langle \Lambda^2 \rangle / E_0^2$, follows from eqs.~(\ref{ave}) and~(\ref{eq:Sigma.full}),
\EQA
\langle \lambda^2 \rangle (c, \v, r^2)
 &\simeq& \left \{ \begin{array}{ll}
 \pl   \lambda^2  \preq(c,r^2) + \v \tau_\eq (c) \dfrac{w(c)}{2}  \left[1 + \O \left (\dfrac{\tau_\eq}{\tau_\sat} \right)\right] ,
 & \macro
 \\ \\
 \langle \lambda \rangle_\eq (c,0) + r^2 \Big [1 - \O \Big (\dfrac{\tau_\sat}{\tau_\eq} \Big ) \Big],
& \micro,  
\end{array} \right. 
\label{lambdastat}
\EEA
where
\EQA
\langle \lambda^2 \rangle_\eq(c,r^2)  & = & \frac{\x}{2c}  +(\langle \lambda^2\rangle_0 + r^2) (1 - \x)^2  
\nonumber \\
& \simeq & \frac{1}{2c} \hspace{1cm} \mbox{for $ c \gg 1$} 
\label{lambdaeq}
\EEA
is the equilibrium average in a fitness landscape. In micro-evolutionary seascapes, this distance remains small, which indicates that the trait distribution~$\W(E)$ efficiently follows the displacements of the fitness peak. In macro-evolutionary seascapes, the mean square distance $\langle \lambda^2 \rangle$ becomes comparable to the driving span~$r^2$; that is, the population no longer follows the moving fitness peak in an efficient way. 

The distribution $Q_\st (\Gamma, E^*)$ describes a stationary state that is manifestly out of equilibrium, i.e., it does not have detailed balance. Its probability  current 
\EQA
\lefteqn{\mathbf J_\st (\Gamma, E^*)  
 =   - \begin{pmatrix} \dfrac{g^{\Gamma \Gamma}}{2N} \dfrac{\partial}{\partial\Gamma} 
-  m^\Gamma  + g^{\Gamma \Gamma} \dfrac{2c}{NE_0^2}  (\Gamma - E^*) 
 \\ 
\v\dfrac{\partial}{\partial E^{*}}  +  \dfrac{\v}{ r^2} (E^* - \cE)
\end{pmatrix} Q_\st (\Gamma, E^*) }
\nonumber  \\  
\nonumber \\ 
& &  \simeq \left \{ \begin{array}{ll}
 \left[- 2\v c  \left ( \begin{matrix}  \hat \Gamma-\hat E^* (1+1/(2 cr^2)) \\ (\hat \Gamma -\hat E^*)
\end{matrix}  \right )  \left(1+  O \left(\dfrac{ \tau_\eq}{\tau_\sat}
\right)\right) \right] Q_\st(\Gamma,E^*) ,
&\macro  \label{def:J}
\\ \\
\left [\dfrac{c \ql \delta \qr}{N} \left (\begin{matrix} \hat E^*  \\ -2 c  r^2 \hat \Gamma\end{matrix} \right ) 
\Big (1 - O \Big ( \dfrac{\tau_\sat}{\tau_\eq}
 \Big) \Big) \right ] Q_\st(\Gamma,E^*) 
&\micro,
\end{array} \right.
\label{J} 
\EEA 
expresses the adaptive motion of the trait mean following the displacements of the 
fitness peak. The probability   current shows a crossover similar to the adaptive part of $\ql \Lambda^2 \qr$ in~(\ref{lambdastat}): it increases linearly for low driving rates and saturates to a constant in the regime of micro-evolutionary fitness fluctuations. 
 
Remarkably, the joint statistics of mean and optimal trait can be associated with evolutionary equilibrium in the limits of low and high driving rates. In the first case, we obtain the equilibrium distribution 
\EQA
Q_\eq (\Gamma, E^*) & = & \lim_{\v \to 0} Q_\st (\Gamma, E^*)
\nonumber \\
& = & \tilde Q_0 (\Gamma) \, \exp[2 N  F_1 (\Gamma | E^*)] \;  R(E^*)
\nonumber  \\  
& = & \frac{1}{Z_\Gamma \sqrt{2 \pi r^2 E_0^2}} \exp \left[ -\frac{2\theta}{\langle \Delta \rangle} (\Gamma-\Gamma_0)^2-\frac{c}{E_0^2} (\Gamma-E^*)^2- \frac{1}{2r^2 E_0^2}(E^*-\mathcal E)^2\right] , 
\label{Qeq}
\EEA
which is the product of a Boltzmann distribution (\ref{QeqGamma}) and a \emph{quenched} weight of the trait optimum $E^*$ given by (\ref{Req}). This distribution satisfies detailed balance; that is, the probability current $\mathbf J_\st (\Gamma, E^*)$ vanishes in the limit $\v \to 0$. In the opposite limit, we obtain the distribution 
\EQA
Q_\infty (\Gamma, E^*) & = & \lim_{\v \to \infty} Q_\st (\Gamma, E^*)
\nonumber \\
& = & \tilde Q_0 (\Gamma) \, \exp[2 N  F_1 (\Gamma | \mathcal E)] \;  R(E^*)
\nonumber  \\  
& = & \frac{1}{Z_\Gamma \sqrt{2 \pi r^2 E_0^2}} \exp \left[ -\frac{2\theta}{\langle \Delta \rangle} (\Gamma-\Gamma_0)^2- \frac{c}{E_0^2} (\Gamma- \mathcal E)^2- \frac{1}{2r^2 E_0^2}(E^*-\mathcal E)^2\right]. 
\label{Qinfty}
\EEA
In this limit, the fast fluctuations of the fitness peak ---~and the associated current  $\mathbf J_\st (\Gamma, E^*)$ given by~(\ref{J}) ---~decouple from the macro-evolutionary dynamics of the mean trait. The latter is governed by the effective fitness landscape
\EQ
F_1 (\Gamma | \mathcal E) = \int  F_1 (\Gamma | E^*) \,  R(E^*) \, dE^* ,
\label{fas}
\EE
which is obtained by averaging over the ensemble (\ref{Req}) of fitness peak positions and it describes stabilizing selection towards the average peak position $\mathcal E$. Accordingly, the scaled average square distance~$\langle \lambda^2 \rangle$, as given by eq.~(\ref{lambdastat}), is the sum of the equilibrium variance $\langle \lambda^2 \rangle_\eq (c,0) $ and the driving span~$r^2$. We can extend the notion of an effective fitness landscape to micro-evolutionary seascapes with a large but finite driving rate ($c \gg 1$, $\v \gg r^2/ \tau_\eq (c)$). Such seascape models still generate stabilizing selection on the trait mean towards the mean peak position $\mathcal E$, but with a reduced effective stabilizing strength  \EQ
c_{\rm eff} \sim c \, \left [1- \frac{2c^2r^2\tau_\sat (\v,r^2)}{\tau_\eq(c)} \right] . 
\label{ceff}
\EE
Similar effective landscapes resulting from micro-evolutionary seascapes have been observed in phenomenological models~\cite{Rivoire2013}.

As shown in Appendix~A, the dynamics of the trait in a punctuated seascape leads to a stationary population ensemble that has the same first and second moments as in the case of a diffusive seascape. 
In particular, the average square displacement between mean and optimal trait, eq.~(\ref{lambdastat}), as well as the averages of divergence, genetic load, and fitness flux described in the following sections coincide for both kinds of seascapes. 

\begin{figure}
\centering
\begin{tabular}{rlll}
&\fig{a} & \fig{b} & \fig{c} \\ \\
 \begin{sideways} \hspace{15pt} centered mean trait, $ \hat \gamma$ \end{sideways}  &
\includegraphics[width=0.3\textwidth]{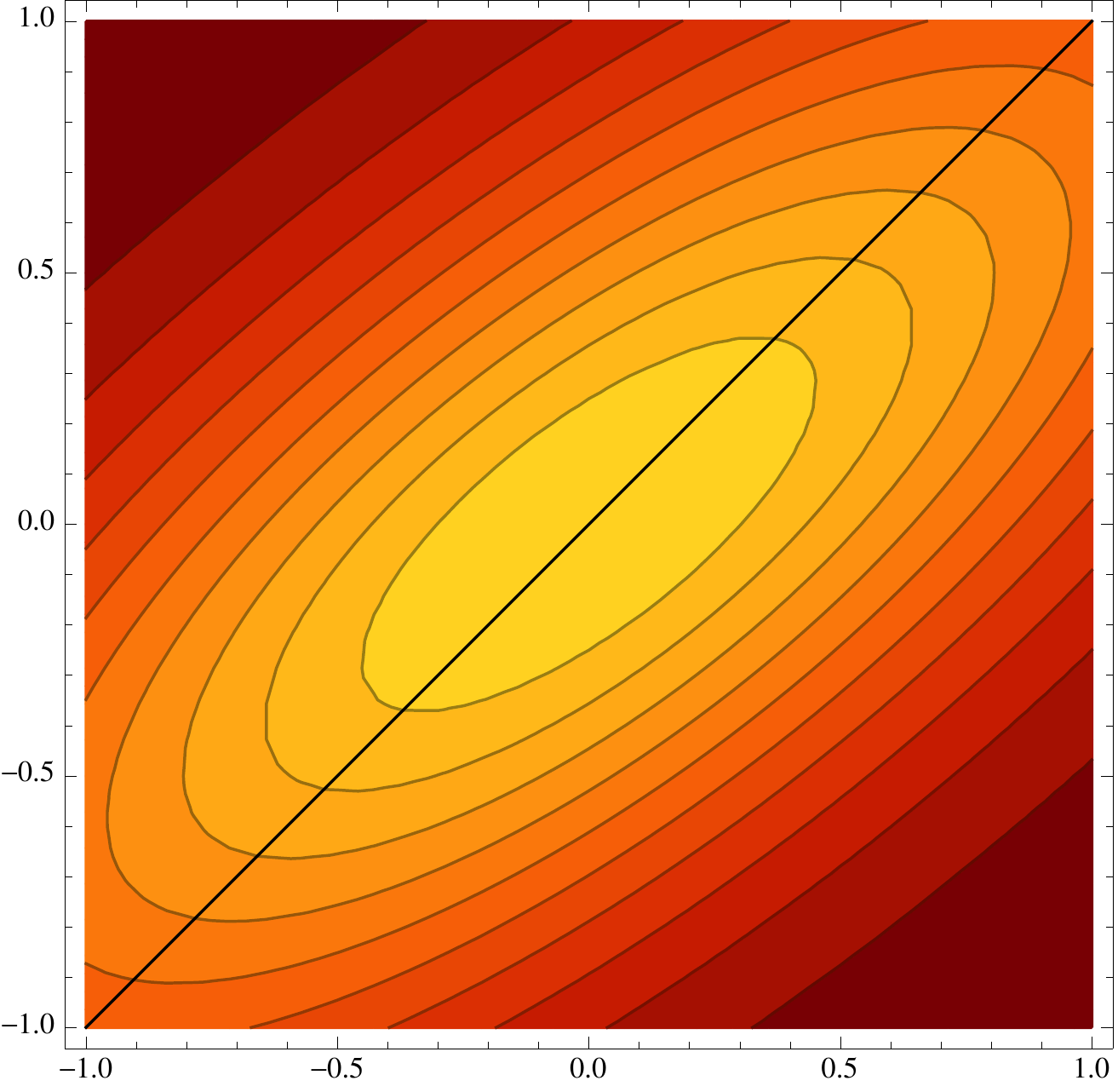}
&\includegraphics[width=0.3\textwidth]{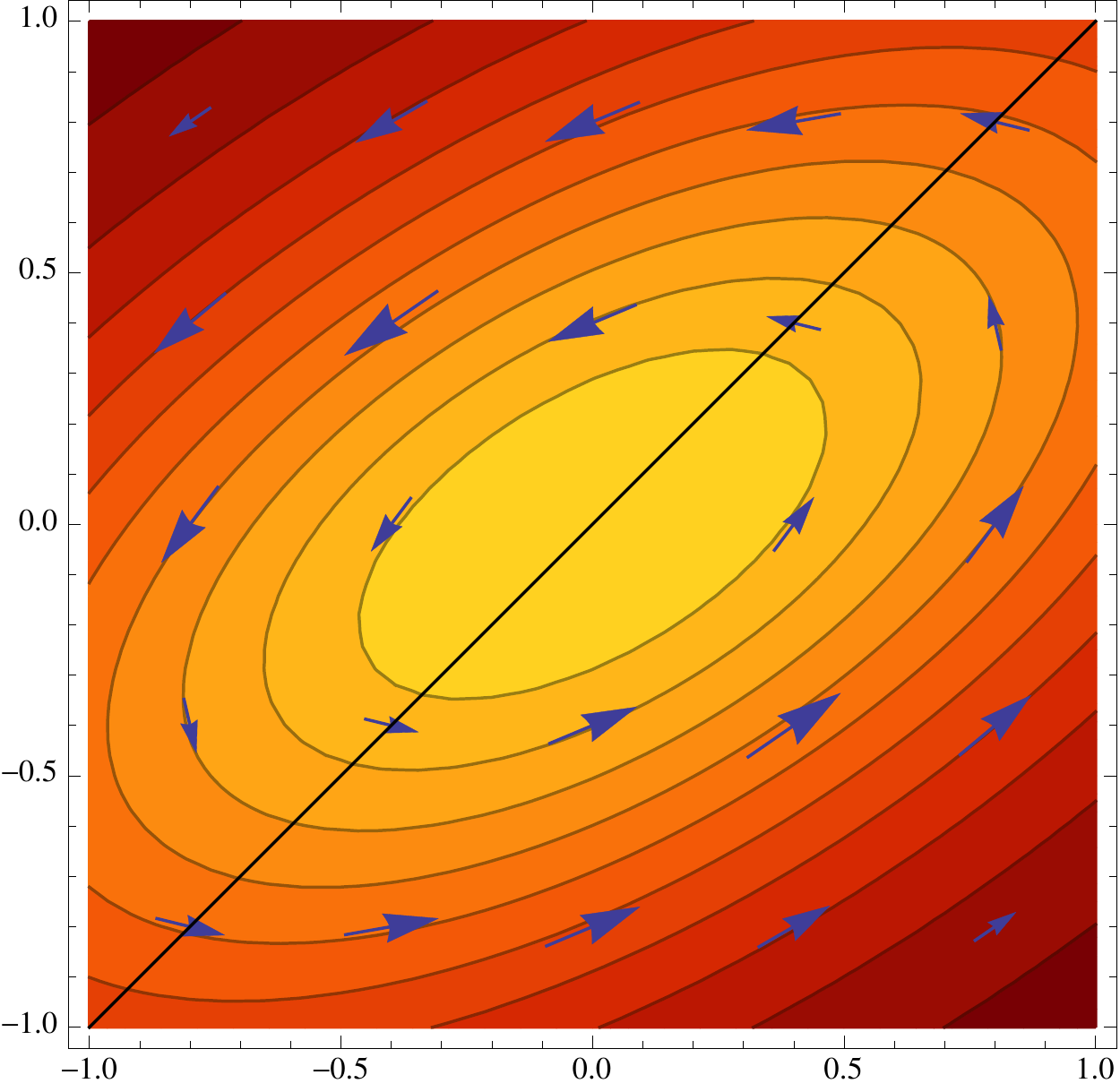} 
&\includegraphics[width=0.3\textwidth]{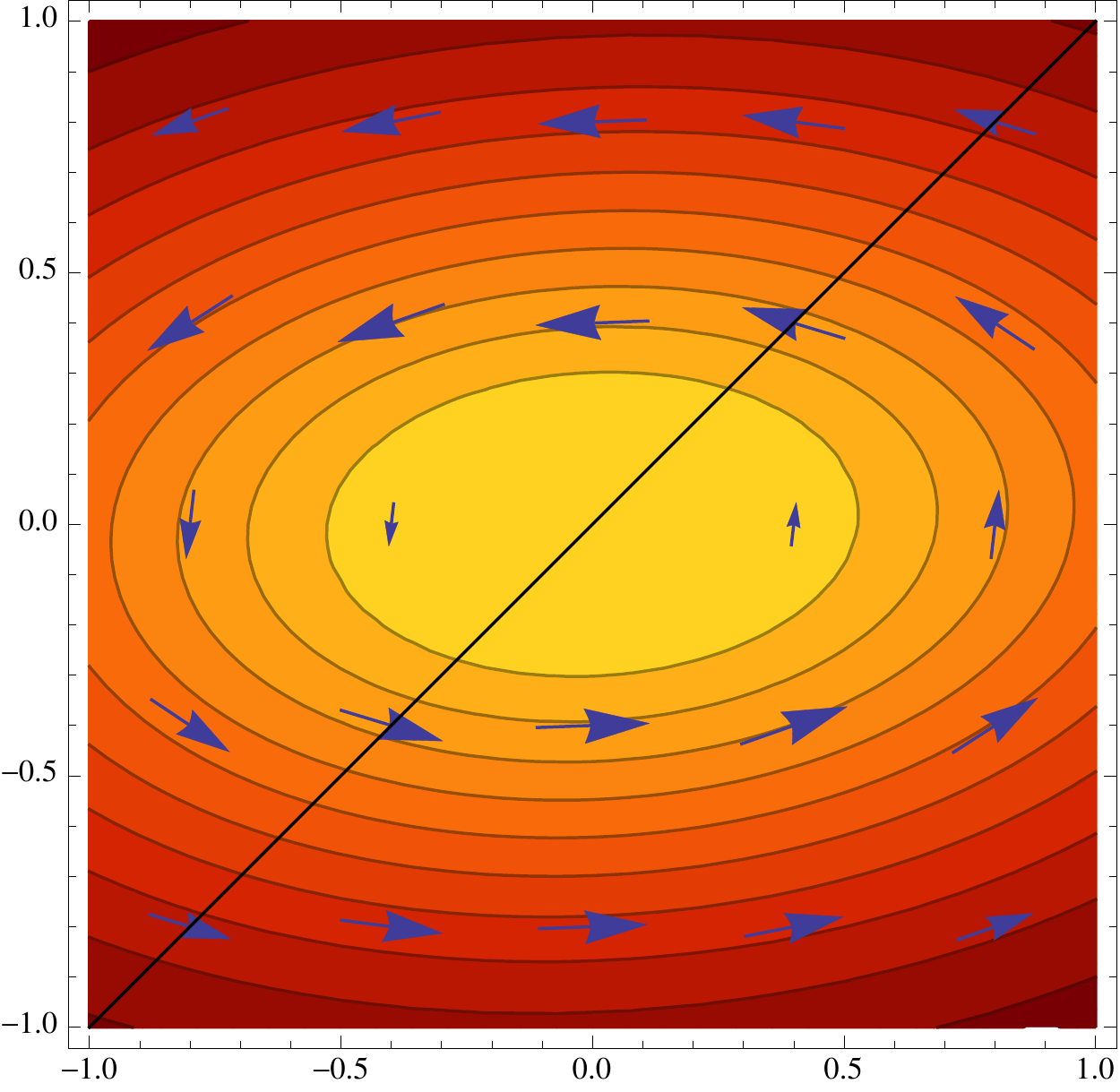} 
 \\ 
& \qquad centered optimal trait, $\hat e^*$ &
\qquad centered optimal trait, $\hat e^*$&
\qquad centered optimal trait, $\hat e^*$ \\ \phantom{.}
\end{tabular}

\begin{tabular}{rll}\\
&\fig{d} & \fig{e} \\  
\begin{sideways} \hspace{25pt}  sq. distance to peak, $\ql \lambda^2 \qr$ \end{sideways} & 
\includegraphics[width=.40\textwidth]{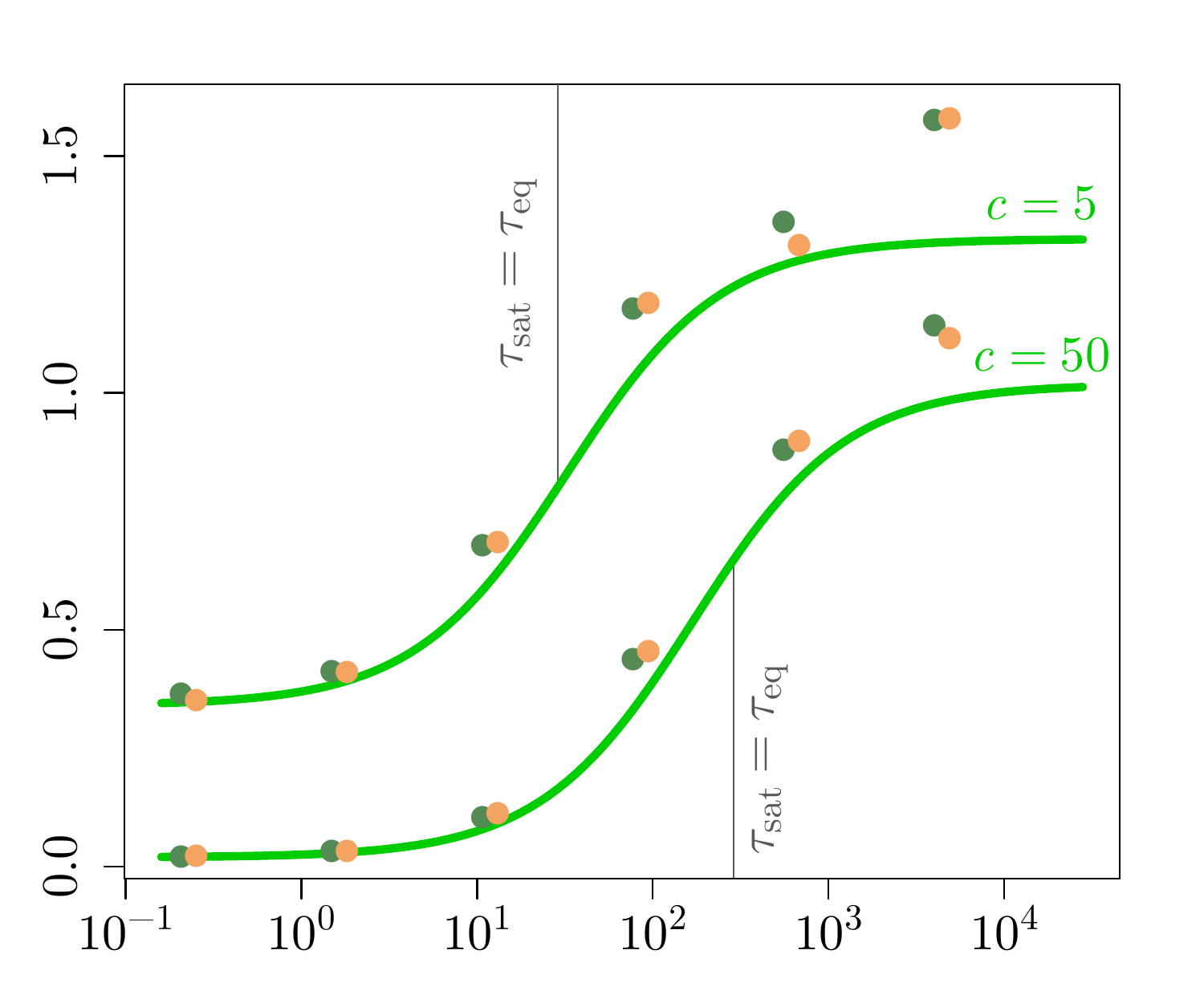}  \hspace*{1cm}&
\includegraphics[width=.40\textwidth]{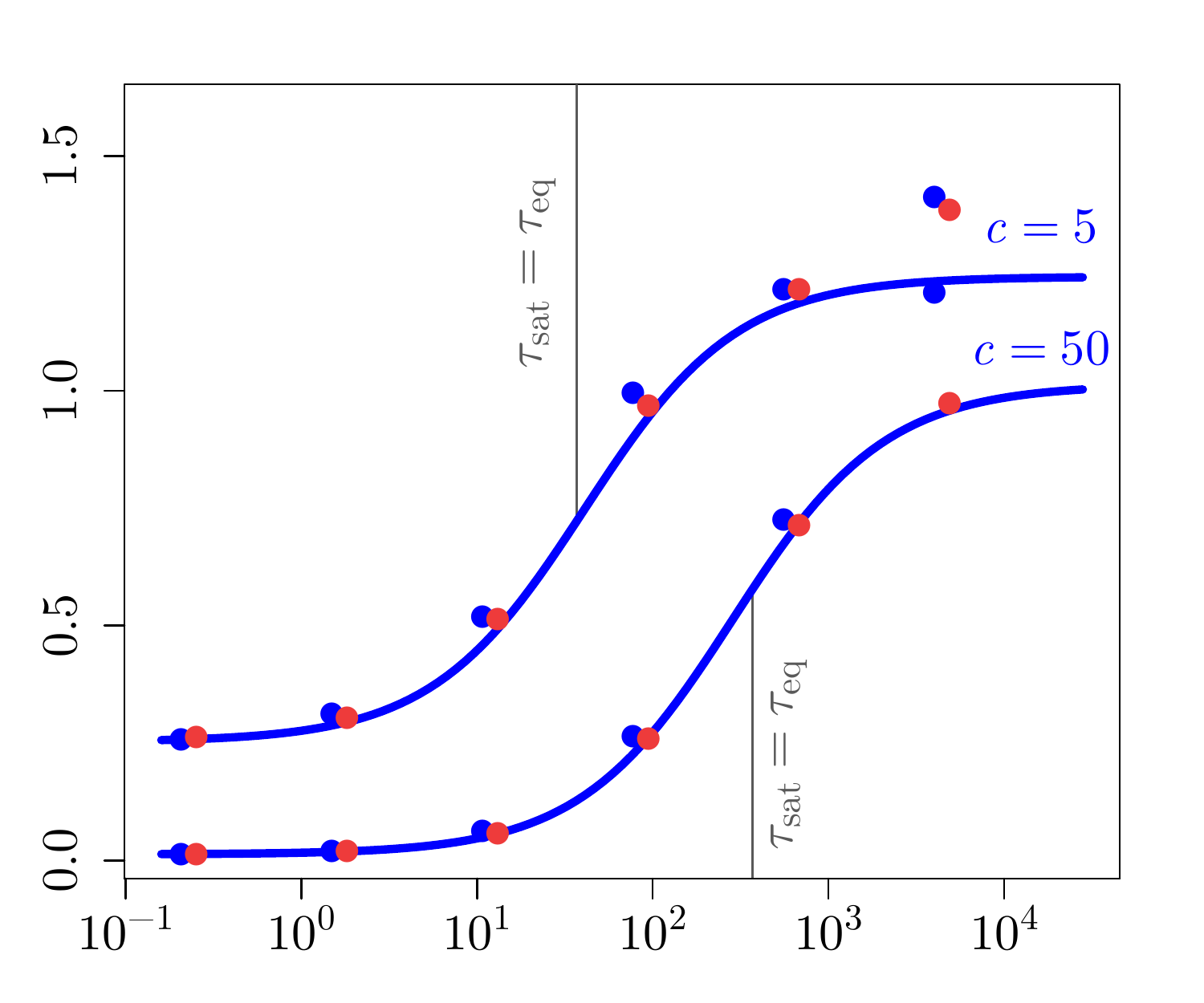} 
\\
&\hspace{50pt} scaled driving rate, $\v /\mu$& \hspace{50pt} scaled driving rate, $\v /\mu$ 
\end{tabular}
\caption{{\bf Stationary distribution of mean and optimal trait in a fitness seascape.} 
({a--c})~The distribution~$Q_\st (\Gamma, E^*)$ is shown 
(a)~in the equilibrium limit ($c=1$, $\v = 0$, $r^2=1$), 
(b)~for an intermediate driving rate ($\v = 1.5 r^2/\tau_\eq$), and 
(c)~in the deep micro-evolutionary regime ($\v = 50 r^2/\tau_\eq$); see eqs.~(\ref{eq:ansatz1}), (\ref{Qeq}), and~(\ref{Qinfty}). The probability current ${\bf J } (\Gamma, E^*)$, which is given by eq.~(\ref{J}), is marked by arrows.
With increasing driving rate, the correlation between $\Gamma$ and $E^*$ is seen to decrease. 
(d,e)~The scaled average square distance between mean and optimal trait value, $\ql\lambda^2\qr$, is plotted against the scaled driving rate $\v /\mu$ for 
(d)~non-recombining and 
(e)~fully recombining populations for different stabilizing strengths $c$. The other parameters are $r^2=1$, $\theta=0.0125$.
This function increases from an equilibrium value for $\v = 0$ to a micro-evolutionary limit value for $\v \to \infty$ with a crossover for $\tau_\sat (\v,r^2) \sim \tau_\eq (c)$, as given by eq.~(\ref{lambdastat}). The analytical results (lines) are compared to simulation results (with parameters $N=100, \ell=100$) for a diffusive seascape (green dots) and for a punctuated seascape (orange dots). 
}
\label{fig:Qstat}
\end{figure}

The properties of the stationary ensemble of mean and optimal trait in a fitness seascape are summarized in Fig.~\ref{fig:Qstat}. 
The stationary distribution  $Q_\st (\Gamma, E^*)$ is shown in Fig.~\ref{fig:Qstat}(a--c) for given parameters $c$, $r^2$, and for different values of the driving rate: in the equilibrium limit ($\v \to 0$), for an intermediate value of $\v$, and in the fast-driving regime ($\v \gg r^2 / \tau_\eq (c)$). The non-equilibrium probability current $\mathbf J_\st (\Gamma, E^*)$ is marked by arrows.  The crossover between micro- and macro-evolutionary fitness seascapes is plotted in Fig.~\ref{fig:Qstat}(d,e) for the scaled  average square distance $\langle \lambda^2 \rangle$ as a function of the driving rate $\v$. Our analytical results are tested by numerical simulations of the underlying Fisher-Wright process \cite{Moehle2001} in a fitness seascape~(\ref{fE}) with diffusive and punctuated peak displacement. The details of the numerical methods for the population simulations are discussed in Appendix~B.

\subsection{Time-dependent trait divergence}
\label{sec:divergence} 
In the previous paper~\cite{Nourmohammad2012}, we have shown that the variance of the trait mean across populations, $\ql \hat \Gamma^2 \qr$, and the average trait diversity $\ql \Delta \qr$ uniquely characterize the stabilizing strength $c$ in a fitness landscape. The ensemble variance $\ql \hat \Gamma^2 \qr$ is just the half of the asymptotic trait divergence $\lim_{\tau\rightarrow \infty}\ql D(\tau)\qr \equiv \lim_{\tau\rightarrow \infty} \ql (\Gamma(t+\tau)-\Gamma(t))^2\qr$.
 As it is clear from the previous subsection, the stationary distribution $Q_\st (\Gamma)$ and its statistics is compatible with different values of the seascape parameters and, hence, cannot uniquely characterize  them. Instead, we will use  the  time-dependent average trait divergence  to characterize the parameters of the fitness seascape. This quantity  can be estimated between   populations in one and two lineages, $\mathcal \ql D^{(\kappa)} \rangle (\tau)$ ($\kappa = 1,2$), as defined in eqs.~(\ref{def:D}) and (\ref{D2}). The average divergence between an ancestral and a descendent population in a single lineage can be written as an expectation value in the stationary ensemble,  
\EQA
 \langle D^{(1)}  \rangle (\tau) &\equiv&  \ql  (\Gamma(t) - \Gamma(t_a))^2 \qr
\nonumber \\
&\equiv& \int \! \d \Gamma \,\d \Gamma_a \,(\Gamma-\Gamma_a)^2 \, \times \left( \raisebox{-11mm}{\includegraphics[height=22mm]{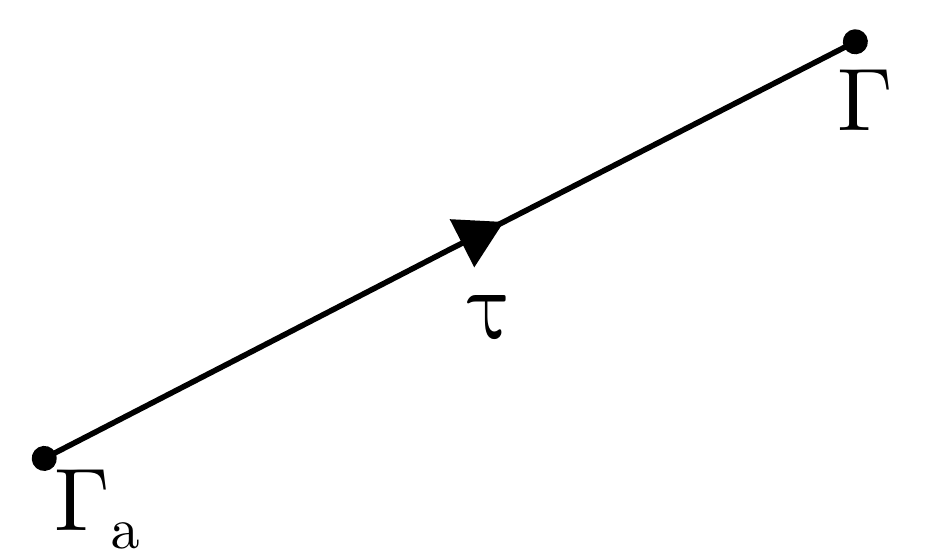} }\right) 
\nonumber \\
 &=&  \int \! \d E^*_a\, \d E^* \,  \d \Gamma_a\, \d\Gamma\, (\Gamma-\Gamma_a)^2 \, G_{\tau}(\Gamma,E^*\, | \, \Gamma_a,E_a^*) \,  Q_\st(\Gamma_a,E_a^*),
\label{def:Divergence} 
\EEA
where $\Gamma_a \equiv \Gamma(t_a)$, $\Gamma \equiv \Gamma(t)$, $E^*_a \equiv E^*(t_a)$, $E^* \equiv E^*(t)$, and $\tau = t - t_a$. The function $G_{\tau}(\Gamma,E^*\, | \, \Gamma_a,E_a^*)$ is the conditional probability (or propagator) for mean and optimal trait values $\Gamma, E^*$ at time $t$, given the values $\Gamma_a, E^*_a$ at time $t_a$. 

In a similar way, the average divergence between two descendent populations evolved from a common ancestor population is given by 
\EQA
\nonumber\ql D^{(2)}  \qr (\tau) &\equiv&  \langle (\Gamma_1(t)-\Gamma_2(t))^2 \rangle_{\substack{\Gamma_1 (t_a) = \Gamma_2 (t_a) \\
 E^*_1 (t_a) = E^*_2 (t_a) }}
\nonumber \\
&\equiv& \int \! \d \Gamma_a \, \d \Gamma_1 \,\d \Gamma_2 \,(\Gamma_1-\Gamma_2)^2 \, \times \left( \raisebox{-11mm}{\includegraphics[height=22mm]{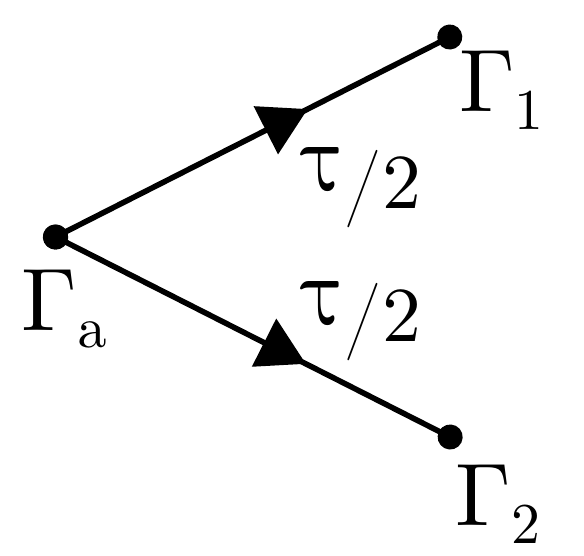} }\right) 
\nonumber \\
&=& \int \! \d E^*_a\, \d E^*_1 \,  \d E^*_2 \,  \d \Gamma_a\, \d\Gamma_1 \, \d\Gamma_2 \; (\Gamma_1-\Gamma_2)^2 \, G_{\tau/2}(\Gamma_1,E_1^*\, | \, \Gamma_a,E_a^*)  \,  
\nonumber \\ 
&& \qquad \qquad\qquad\qquad\qquad\qquad \times G_{\tau/2}(\Gamma_2,E_2^*\, | \, \Gamma_a,E_a^*) \,  Q_\st(\Gamma_a,E_a^*),  \label{def:Divergence2} 
\EEA
where $\Gamma_a \equiv \Gamma_1(t_a) = \Gamma_2 (t_a)$, $E^*_a \equiv E_1^*(t_a)=E_2^*(t_a)$, $\Gamma_i \equiv \Gamma_i (t)$, $E^*_i \equiv E^*_i (t)$ ($i = 1,2$), and $\tau \equiv 2(t - t_a)$. 
In Appendix~A, we use the  Langevin formalism to derive analytical expressions for these propagators and  the resulting scaled divergence $\ql d^{(\kappa)} \rangle (\tau)  \equiv \ql D^{(\kappa)} \rangle (\tau) / E_0^2,\; (\kappa=1,2)$. We obtain
\EQA
\lefteqn{\ql d^{(\kappa)} \qr (\tau; c, \v, r^2) = \frac{\tau_\eq (c)}{\tilde \tau (c)} \Big [1-e^{- \tau / \tau_\eq (c)} \Big ] }
\nonumber \\
&&+ \v \, \xv \xmv 
\left [ \tTwo \big (1- \e^{-\tau/\tTwo} \big ) - \tOne \left (1- \e^{-\tau/\tOne} \right )   \right ]  \nonumber 
\\
& & - 2 (\kappa -1) \frac{\v}{\tau_\eq^{-1}(c)+\tau_\sat^{-1}(v,r^2)}\, \xmv^2  \left [ \e^{- \tau/(2\tTwo)}-\e^{-\tau/(2\tOne)} \right ]^2 .
\label{dkappa}
\EEA 
The difference between the two divergence measures is a consequence of the non-equilibrium adaptive dynamics, which violate  detailed balance.  Equation (\ref{dkappa}) is valid for diffusive and for punctuated fitness seascapes. It contains the three characteristic time scales defined in the previous section: the {\em drift time} $\tilde \tau (c)$ is the scale over which  the diffusion of the trait mean, in the absence of any fitness seascape, generates a trait divergence of the order of the neutral trait span $E_0^2$; the {\em equilibration time} $\tau_\eq (c)$ governs the relaxation of the population ensemble to a mutation-selection-drift equilibrium in a fitness landscape of stabilizing strength $c$; the {\em saturation time} $\tau_{\rm sat}$ is defined by the mean square displacement of the fitness peak reaching the driving span $r^2$. 
Here, we focus on  fitness seascapes with substantial stabilizing strength and with a driving span of order of the neutral trait span ($c \gtrsim 1$,  $r^2 \sim 1$). This selection scenario is biologically relevant: it describes adaptive processes that build up large trait differences by continuous diffusion or recurrent jumps of the fitness peak. 

In macro-evolutionary seascapes, the equilibration time and the non-equilibrium saturation time are well-separated, $\tau_\eq (c)\ll \tau_{\rm sat} (\v,r^2$). This results in three temporal regimes of the trait divergence:
\begin{itemize}
\item {\bf Drift regime,  $\tau \lesssim \tOne$.} The scaled divergence takes the form 
\EQA
\ql d^{(\kappa)} \qr( \tau) & = & 2  \pl  \hat \gamma^2  \pr_\eq  \left(1-e^{-\tau/\tOneA}\right)  
\nonumber \\
& \simeq & \frac{\langle \delta \rangle}{2N}\,\tau \,\big(1  + O(\tau/ \tau_\eq) \big) 
\label{eq:div_short}
\qquad (\kappa=1,2),
\EEA
with an initial increase due to genetic drift and relaxation to an equilibrium value 
\EQ
2 \pl  \hat \gamma^2 \rangle_\eq (c) = \frac{w(c)}{c}
\EE
due to stabilizing selection. 

\item{\bf Adaptive regime, $\tOne \lesssim \tau \ll  \tTwo$}. The scaled trait divergence follows 
\EQA
\ql d^{(\kappa)}\qr( \tau) & = &\left[ 2 \pl  \hat \gamma^2  \preq \left(1- \v \,\tilde \tau\,\kappa\,\x^2\right) 
+\v\,\x^2 \,\tau \right]\left[1+ O\left(e^{-\tau/\tOneA},{\tau}/{ \tTwoA}\right) \right]
\nonumber \\ \nonumber &&\hspace{0.65\textwidth} 
\nonumber \\
& \simeq & \left[ 2 \pl  \hat \gamma^2  \preq +\v\,(\tau -\kappa \tau_\eq (c))\right] \left[ 1+\; O \left((\theta c)^2, e^{-\tau/\tOneA},\tau/ \tTwoA\right) \right] 
\;\;\; (\kappa=1,2)
\EEA
In this regime, the trait divergence is the sum of an (asymptotically constant) equilibrium component and an adaptive component, which increases with slope $\v$. In a macro-evolutionary fitness seascape, this slope is, by definition, smaller than the slope in the initial drift regime (\ref{eq:div_short}), which allows for a clear delineation of the two regimes in empirical data. This feature will be exploited in our selection test for quantitative traits, which will be discussed in section~5. 

\item{\bf Saturation regime, $ \tau\gtrsim \tTwo$}. On the largest time scales, the divergence 
\EQ
\ql d^{(\kappa)}\qr( \tau) \approx 2\ql \hat \gamma^2\qr_\eq+2r^2\x\xv \big ( 1-e^{- \tau/\tTwoA} \big )  
\qquad (\kappa=1,2)
\EE
approaches its non-equilibrium saturation value
\EQ
\ql \hat \gamma^2 \qr_{\st} (c, \v, r^2) =  2\ql \hat \gamma^2\qr_\eq+2r^2 \x\xv ,
\EE 
which equals the $\Gamma$-variance of the stationary distribution $Q_\st (\Gamma, E^*)$, and is primarily  determined 
by the driving span $r^2$. In empirical data, this regime is often well beyond the depth of the phylogeny and, hence, not observable. 
\end{itemize}

In micro-evolutionary seascapes, the saturation of fitness fluctuations occurs faster than the equilibration of the trait under stabilizing selection, i.e., $\tau_{\rm sat} (\v, r^2) \lesssim \tau_\eq (c)$. Hence, there is a direct crossover from the drift regime to the saturation regime. For fast micro-evolutionary fitness fluctuations, $\tau_{\rm sat} (\v, r^2) \ll \tau_\eq (c)$, the constraint on the trait equals that in an effective fitness landscape with stabilizing strength $c_{\rm eff}$ given by eq.~(\ref{ceff}). In this regime, time-dependent trait divergence data alone can no longer resolve adaptive evolution in a fitness seascape from equilibrium in the corresponding effective fitness landscape; this requires additional information on the trait diversity.  

\begin{figure}
\centering 
\begin{tabular}{rll}
 & \fig{a}  & \fig{b}  \\
\begin{sideways} \hspace{30pt} 
  scaled  trait divergence, $\ql d^{(1)} \qr(\tau)$\end{sideways} &
\includegraphics[width=0.4\textwidth]{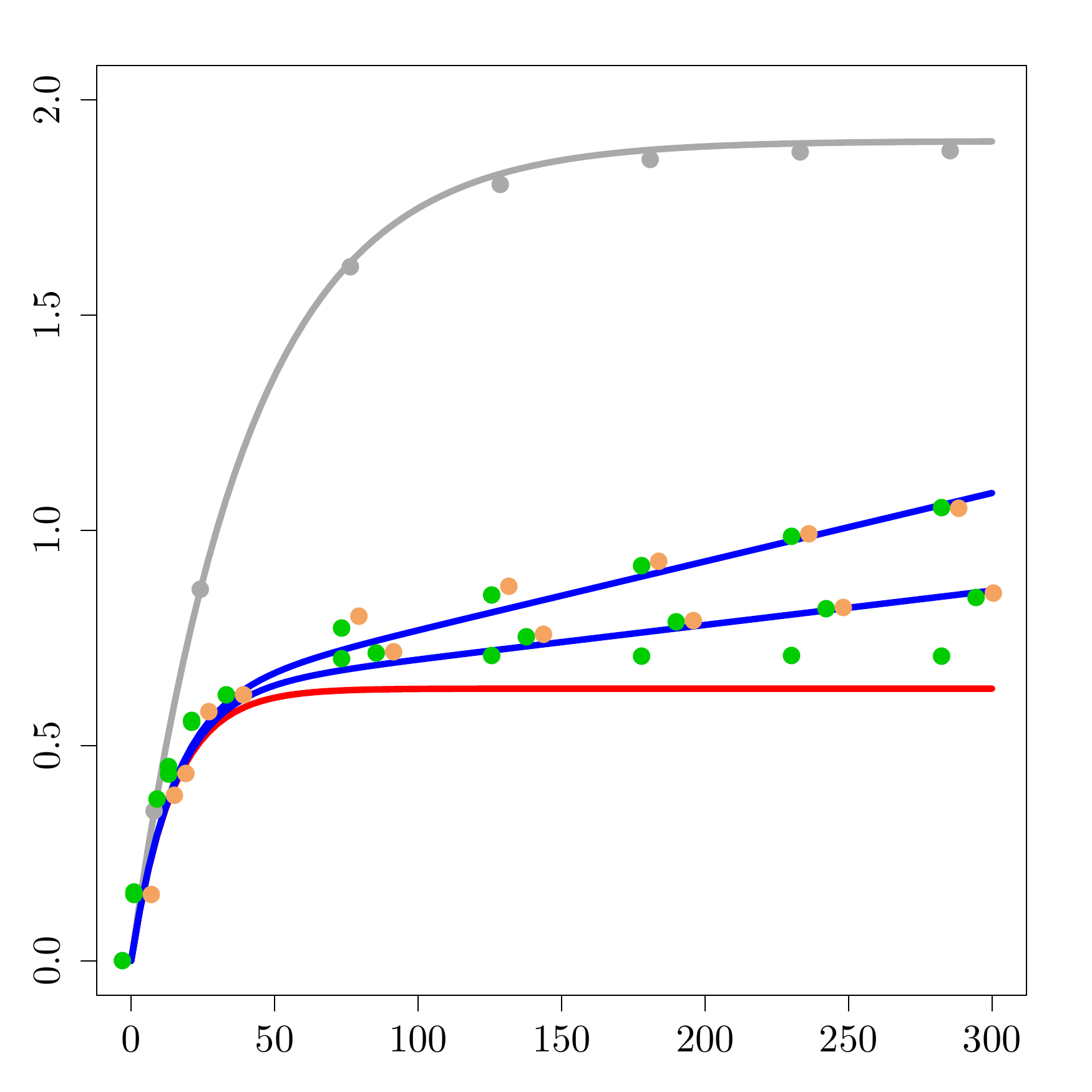} &
\includegraphics[width=.4\textwidth]{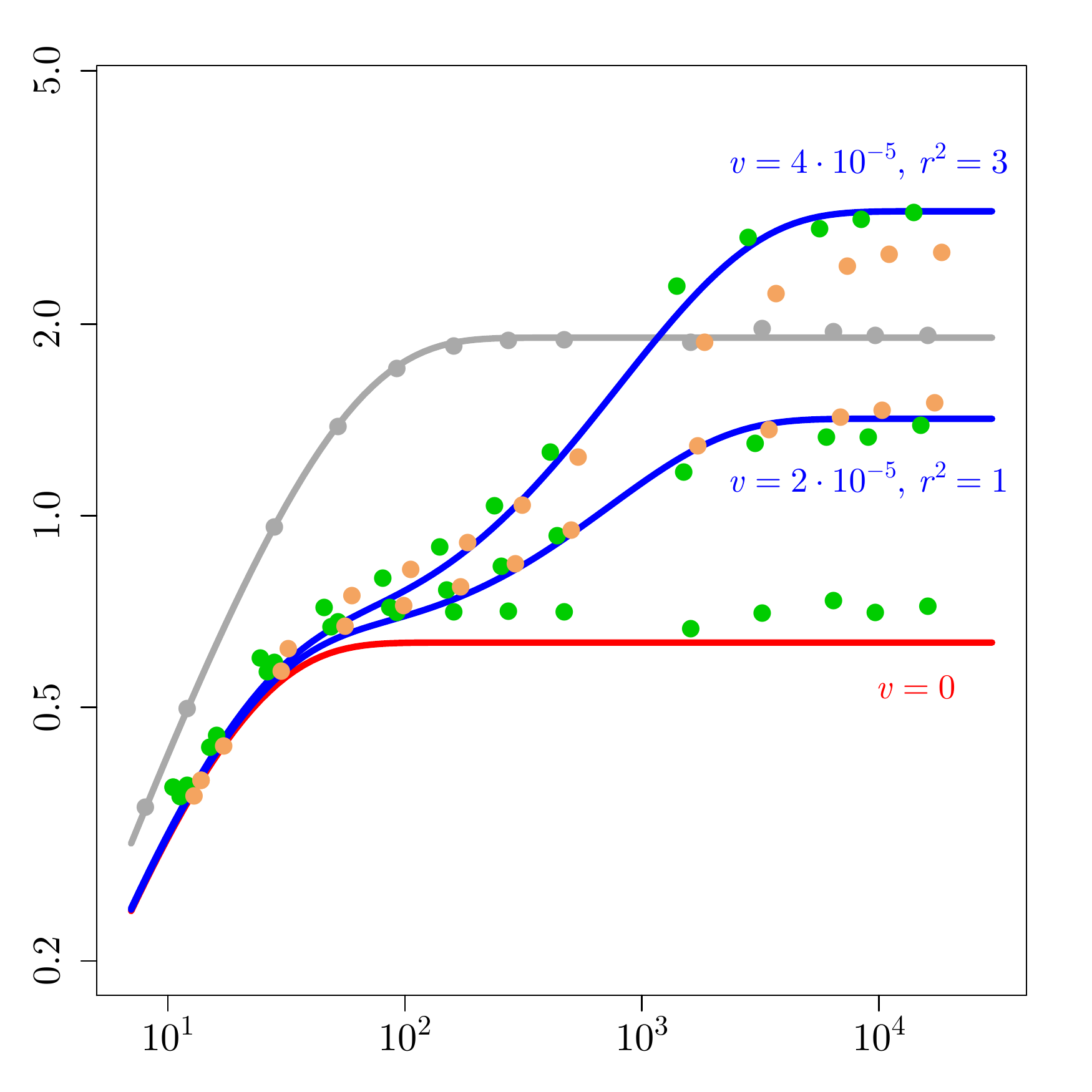} \\  
& \hspace{30pt} scaled divergence time, $\tau/N$ &
\hspace{30pt} scaled divergence time, $\tau/N$ \\
& \fig{c}   \\
\begin{sideways} \hspace{30pt} 
  scaled  trait divergence, $\ql d^{(1)} \qr(\tau)$\end{sideways} &
 \includegraphics[width=0.4\textwidth]{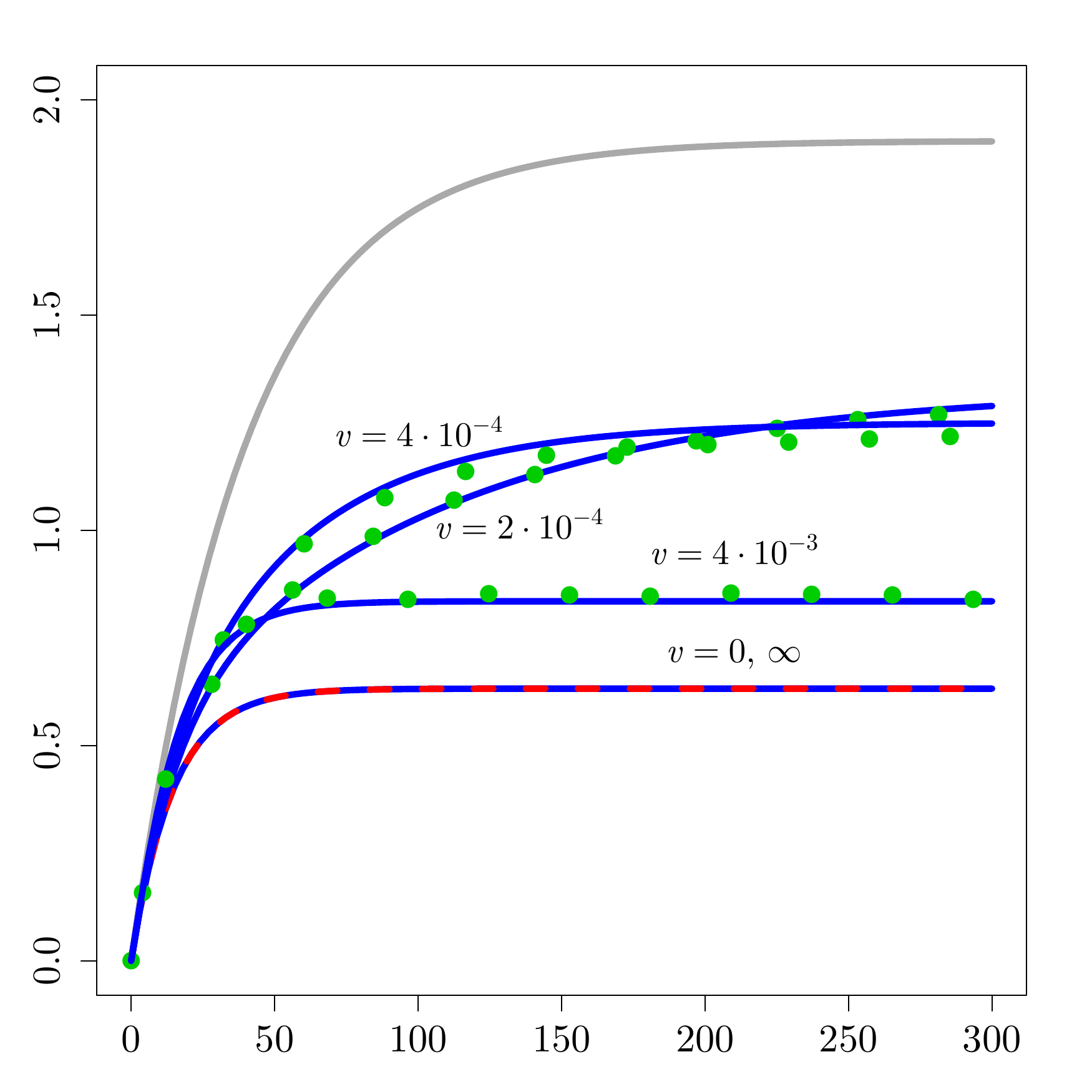} 
\\
&
\hspace{30pt} scaled divergence time, $\tau/N$ \\
\end{tabular}
\caption{{\bf Time-dependence of the trait divergence.}
The scaled average divergence $\langle d^{(1)} \rangle(\tau)$ is shown as a function of the scaled divergence time $\tau/N$ for  three cases: neutral evolution ($c = 0$; grey lines), conservation in a static fitness landscape ($c = 1$, $\v = 0$; red line), and adaptation in a macro-evolutionary fitness seascape ($c = 1$, $\v > 0$; blue lines). Other parameters: $\theta=0.0125,\,N=100,\, \ell=100,\,\mathcal E=0.7\ell$. The analytical results of eq.~(\ref{dkappa})  (lines) are compared to simulation results for asexual evolution in diffusive and punctuated fitness seascapes (green and orange dots, respectively). The corresponding results for fully recombining genomes are shown in Fig.~\ref{fig:freerec}. 
(a,b)~Macro-evolutionary seascapes, $\tau_\sat (\v,r^2) > \tau_\eq (c)$. (a) Linear plot: For $\tau \lesssim \tau_\eq$, the trait evolution is dominated by genetic drift and is independent of selection.   For $\tau \gtrsim \tau_\eq$, the seascape data show an adaptive divergence component proportional to $\v \tau$; the landscape data saturate to an equilibrium divergence set by stabilizing selection. 
(b)~Logarithmic plot: This also shows the non-equilibrium saturation of the seascape data on time scales $\tau \sim \tau_{\rm sat}$, when the divergence reaches twice the driving span~$r^2$. 
(c{)}~Micro-evolutionary seascapes $\tau_\sat (\v,r^2) < \tau_\eq (c)$. There is a single cross-over from the drift regime  for smaller values of $\tau$ to the saturation regime for larger values of $\tau$. The divergence $\langle d^{(1)} \rangle(\tau)$ equals that in an effective fitness landscape of stabilizing strength $c_{\rm eff} < c$. The limit $\v \to \infty$ has $c_{\rm eff} = c$; i.e., the function $\langle d^{(1)} \rangle(\tau)$ becomes identical to the case $\v = 0$ (blue--red dashed line). 
}
\label{fig:acfoverDlong}
\end{figure}

Fig.~\ref{fig:acfoverDlong} and Fig.~\ref{fig:freerec}(a--b) show the scaled  divergence $\ql d^{(1)} \qr(\tau)$ for selection parameters $c$ and $\v$ covering macro-evolutionary and micro-evolutionary fitness seascapes. The analytical expression of eq.~(\ref{dkappa}) is seen to be in good agreement with numerical simulations for diffusive and punctuated fitness fluctuations. 

\subsection{Stationary trait diversity}
\label{chap:delta}

As discussed in section~2, our diffusion theory predicts that the movements of the  optimum trait in  a single-peak fitness seascape of the form (\ref{fE}) only affects the evolution of the  trait mean in the population and  not  the trait diversity. The statistics of the trait diversity remains similar to the case of evolution under stabilizing selection, which is characterized by a time-invariant fitness function,  $F_2( \Delta) = - c_0 \, \Delta$. The resulting equilibrium distribution $Q_\eq (\Delta)$ is the product of the neutral mutation-drift equilibrium $Q_0 (\Delta)$, which is given in eqs.~(53) and~(55) of \cite{Nourmohammad2012} and a Boltzmann factor  from the scaled fitness landscape, $Q_\eq (\Delta) = Q_0 (\Delta) \, \exp[- c_0 \, \Delta]$. These distributions determine the average diversity 
\EQ
\langle \Delta \rangle \equiv  \int \! \d \Delta\, \Delta \, Q_\eq (\Delta)
\EE
and its neutral counterpart $\langle \Delta \rangle_0$, as well as the scaled expectation values $\langle \delta \rangle \equiv \langle \Delta \rangle / E_0^2$ and $\langle \delta \rangle_0 \equiv \langle \Delta \rangle_0 / E_0^2$. The selective constraint on the trait diversity enters the diffusion coefficient of the trait mean in eq.~(\ref{diffGamma}), which sets the drift  time scale $\tilde \tau (c) = (1/2\mu) (\langle \delta \rangle_0 / \langle \delta \rangle (c))$, as 
given by eq.~(\ref{tautilde}). 
The distributions $Q_0 (\Delta)$ and $Q_\eq(\Delta)$ can be written in closed analytical form; unlike in the case of the trait mean, these distributions depend directly on the rate of recombination in the population~\cite{Nourmohammad2012}. We obtain the scaled neutral expectation value $\langle \delta \rangle_0 = 4 \theta (1 - 4 \theta + O(\theta^2))$, which is independent of the recombination rate, and the selective constraint
\EQ
\frac{\langle \delta \rangle (c)}{\langle \delta \rangle_0} = 
\left \{ \begin{array}{ll}
1 - 4 \theta c + \O((\theta c)^2)& \mbox{ for $\theta c \ll 1$,} 
\\
(4 \theta c)^{-1/2} + \O((\theta c)^{-1}) & \mbox{ for $\theta c \gg 1$}
\end{array} \right. 
\label{deltaconstraint}
\EE
in non-recombining populations. We note that this constraint depends only on the product $\theta c$; therefore, it remains weak over a wide range of parameters ($c \lesssim 1/\theta$), which includes strong selection effects on the trait mean~\cite{Nourmohammad2012}. The full crossover function and the corresponding expressions for fully recombining populations are given in eqs.~(68)~--~(73) of ref.~\cite{Nourmohammad2012}.

\begin{figure}
\centering 
\begin{tabular}{rll} &\fig{a}&\fig{b}\\
\begin{sideways} \hspace{25pt} scaled trait diversity, $\ql \delta\qr/\ql \delta  \qr_0$ \end{sideways} &
\includegraphics[width=.40\textwidth]{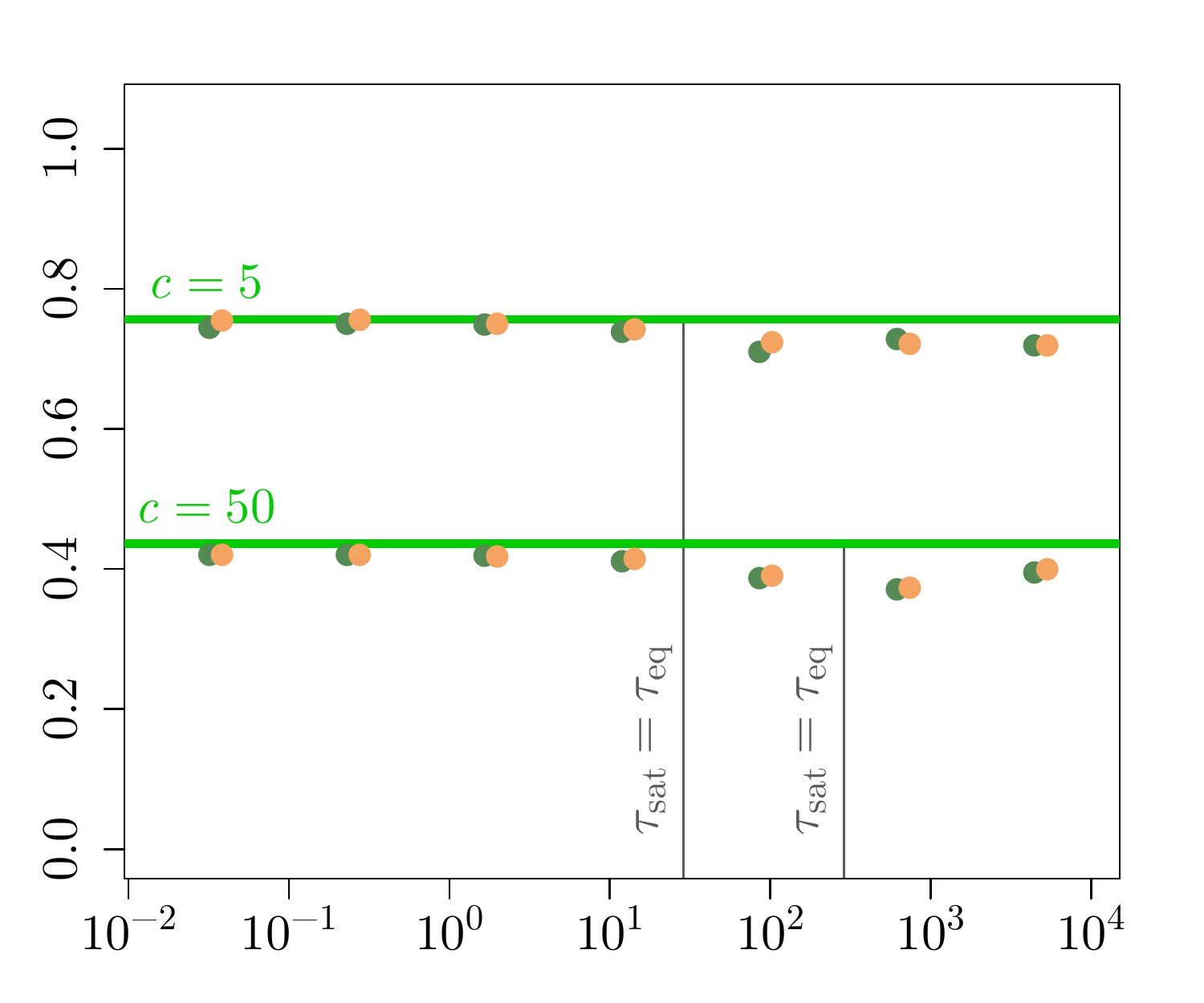} \hspace*{1cm}&
\includegraphics[width=.40\textwidth]{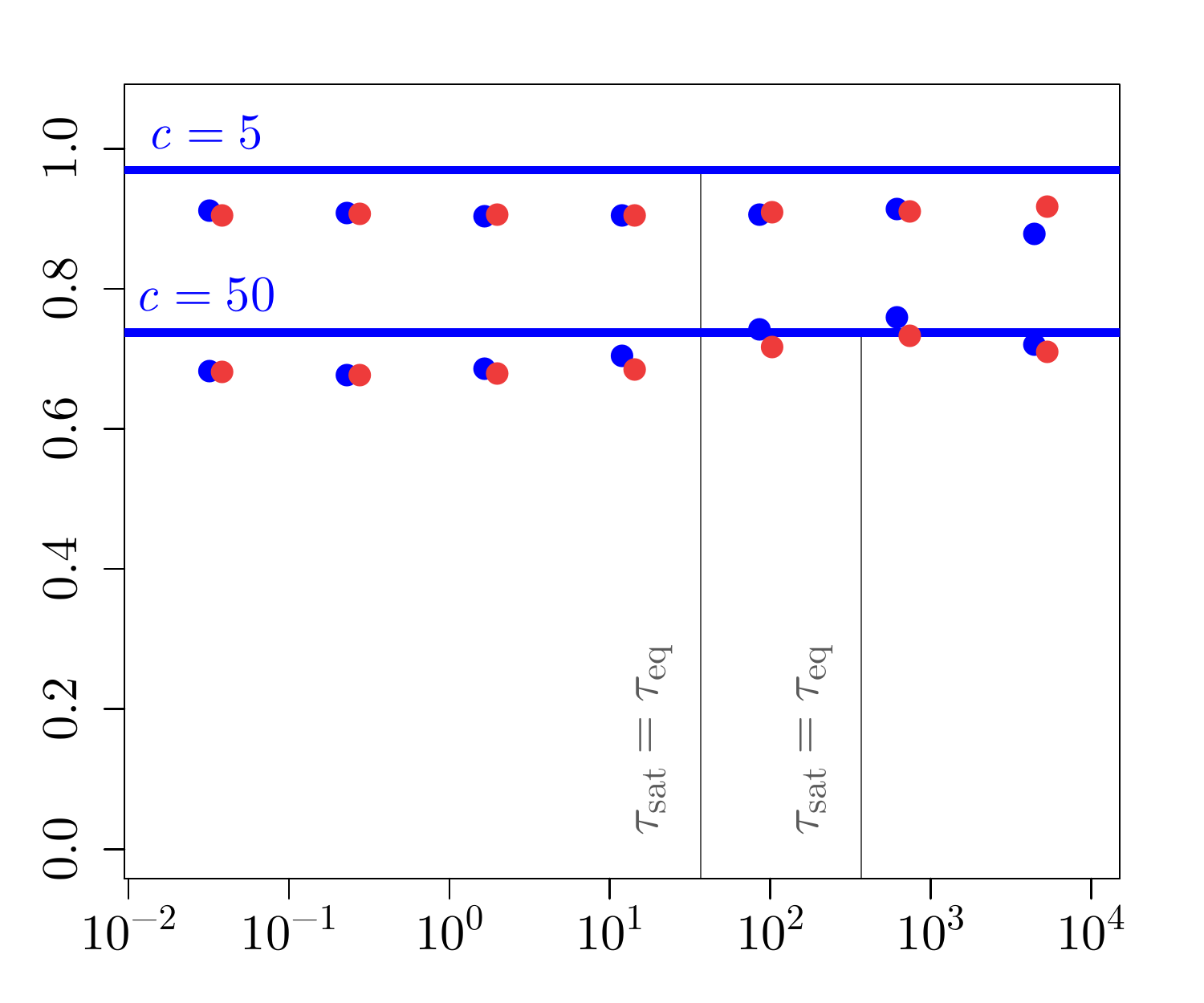}\\ 
&  \hspace{50pt} scaled driving rate, $\v / \mu$ &  \hspace{50pt} scaled driving rate, $\v / \mu$
\end{tabular}
\caption{{\bf Stationary trait diversity}.
The figure shows the average trait diversity $\ql \delta \qr $ (in units of the neutral average $\ql \delta \qr_0$) in a fitness seascape as a function of the scaled driving rate $\v /\mu$ for different values of the stabilizing strength ($c = 5, 50$, top to bottom); other parameters are as in Fig.~\ref{fig:acfoverDlong}. The equilibrium predictions of diffusion theory (lines), which do not depend on $\v$, are compared to simulation results of the adaptive process of 
(a)~non-recombining and 
(b)~fully recombining populations in diffusive (green/orange dots) and punctuated seascapes (blue/red dots). The simulation results confirm evolutionary equilibrium of the trait diversity.  }
\label{Delta}
\end{figure}

The numerical simulations reported in Fig.~\ref{Delta} show that the average diversity in diffusive fitness seascapes is well represented by the equilibrium value throughout the crossover from macro- to micro-evolutionary driving rates, and over a wide range of stabilizing strengths. Theoretically, the results of the diffusion theory are valid for adaptive processes unless recurrent selective sweeps reduce the trait diversity within the population. Such sweeps are more prominent in punctuated fitness seascapes due to  sudden changes of the trait optimum. We expect a significant reduction  in  trait diversity due to the large and frequent jumps of the trait optimum  in  fitness seascapes with very strong stabilizing selection. This regime is beyond the scopes of this paper.

\section{Fitness and entropy of adaptive processes}
\label{sec:Fitness}

The distributions of the trait mean and diversity  determine the fitness statistics of an ensemble of populations in the stationary state. These statistics can quantify the cost and the amount of adaption for the evolution of molecular traits. We also evaluate the predictability of the trait evolution in an ensemble of populations after diverging  from a common ancestral population.

\subsection{Genetic load} 
\label{sec:Load}

The genetic load of an individual population is defined as the difference between the maximum fitness and the mean fitness \cite{Crow:1958,Crow:1965,Haldane:1957,Muller:1950}, 
\EQ
L(t)   \equiv   f^* - \ol  f(t).
\EE
For a quantitative trait in a quadratic fitness seascape of the form (\ref{fE}), we can decompose the load into contributions of the trait mean and diversity,  
\EQ
L(t)  =  f^* - c_0 \big (\Gamma(t)  - E^*(t) \big )^2 - 2c_0 \Delta (t).   \label{load}
\EE
In the stationary population ensemble (\ref{eq:ansatz1}), the average scaled genetic load can be written as the sum of an equilibrium and an adaptive component, 
\EQA
\langle 2 N L \rangle (c, \v, r^2) 
& = & c \big [ \langle \lambda^2 \rangle (c, \v, r^2) + \langle \delta \rangle  (c) \big ] 
\nonumber \\ 
& = & c \big [ \langle \lambda^2 \rangle_\eq (c, r^2) + \langle \delta \rangle  (c) \big ] 
+ c \big [ \langle \lambda^2 \rangle (c, \v, r^2) - \langle \lambda^2 \rangle_\eq (c, r^2) \big ]
\nonumber \\
& \equiv & 2N L_\eq (c, r^2) + 2N L_\ad (c, \v, r^2);  \label{eq:fullload}
\EEA
these components can be computed analytically from eqs.~(\ref{lambdastat}) and (\ref{lambdaeq}). A simple form is obtained for fitness seascapes of substantial stabilizing strength ($c \gtrsim 1$), 
\EQA
2N L_\eq & \simeq & \frac{1}{2} + \O(1/c, \theta c),
\\ 
\nonumber \\
2N L_\ad (c, \v, r^2) & \simeq  &
\left \{
\begin{array}{cc}
\v \, \tilde \tau (c)  \left[1 + \O \left (\dfrac{\tau_\eq}{ \tau_\sat} \right )\right], & \macro
\\ \\
c r^2 \Big [1 - \O\Big (\dfrac{\tau_\sat}{\tau_\eq}\Big ) \Big ],   & \micro,
\end{array} \right.  \label{adaptiveload}
\EEA
where the drift scale $\tilde \tau (c)$ is given by eqs.~(\ref{tautilde}) and~(\ref{deltaconstraint}). From these expressions, we read off three relevant properties of the genetic load. 

First, the equilibrium load depends on $c$ only via its diversity component; this dependence remains weak even for substantial stabilizing selection ($1 \lesssim c \lesssim 1/\theta$). The equilibrium load component related to the trait mean, $c \langle \lambda^2 \rangle_\eq (c)$, becomes universal in this regime: the fluctuations of $\Gamma$ are constrained to a fitness range of order $2N L_\eq \simeq 1/2$ around $E^*$, irrespectively of the stabilizing strength and the molecular details of the trait~\cite{Nourmohammad2013}. For a $d$-component trait as in Fisher's geometrical model \cite{Fisher:1930wy},  this formula generalizes to $2N L_\eq \simeq d/2$ \cite{Hartl1998,Poon:2000vg,Tenaillon:2007jd}. This is  a direct evolutionary analogue of the equipartition
theorem in statistical thermodynamics, which states that every degree
of freedom that enters the energy function quadratically contributes an
average of $k_BT/2$ to the total energy of a system at temperature $T$ (the
proportionality factor~$k_B$ is Boltzmann's constant).~\cite{Nourmohammad2013}.

Similarly, the adaptive load component depends only weakly on $c$, via the drift scale $\tilde \tau (c)$. At a fixed value of $\Gamma$, the stochastic displacement of the fitness peak induces a fitness cost proportional to $c$; however, this effect is largely offset by an adaptive response that becomes faster with increasing $c$.  

\begin{figure}[t]
\centering
\begin{tabular}{rlrl}
&\fig{a} & & \fig{b}  \\
\begin{sideways} \hspace{40pt} scaled genetic load, $\ql 2N L \qr$ \end{sideways} &
\includegraphics[width=.40\textwidth]{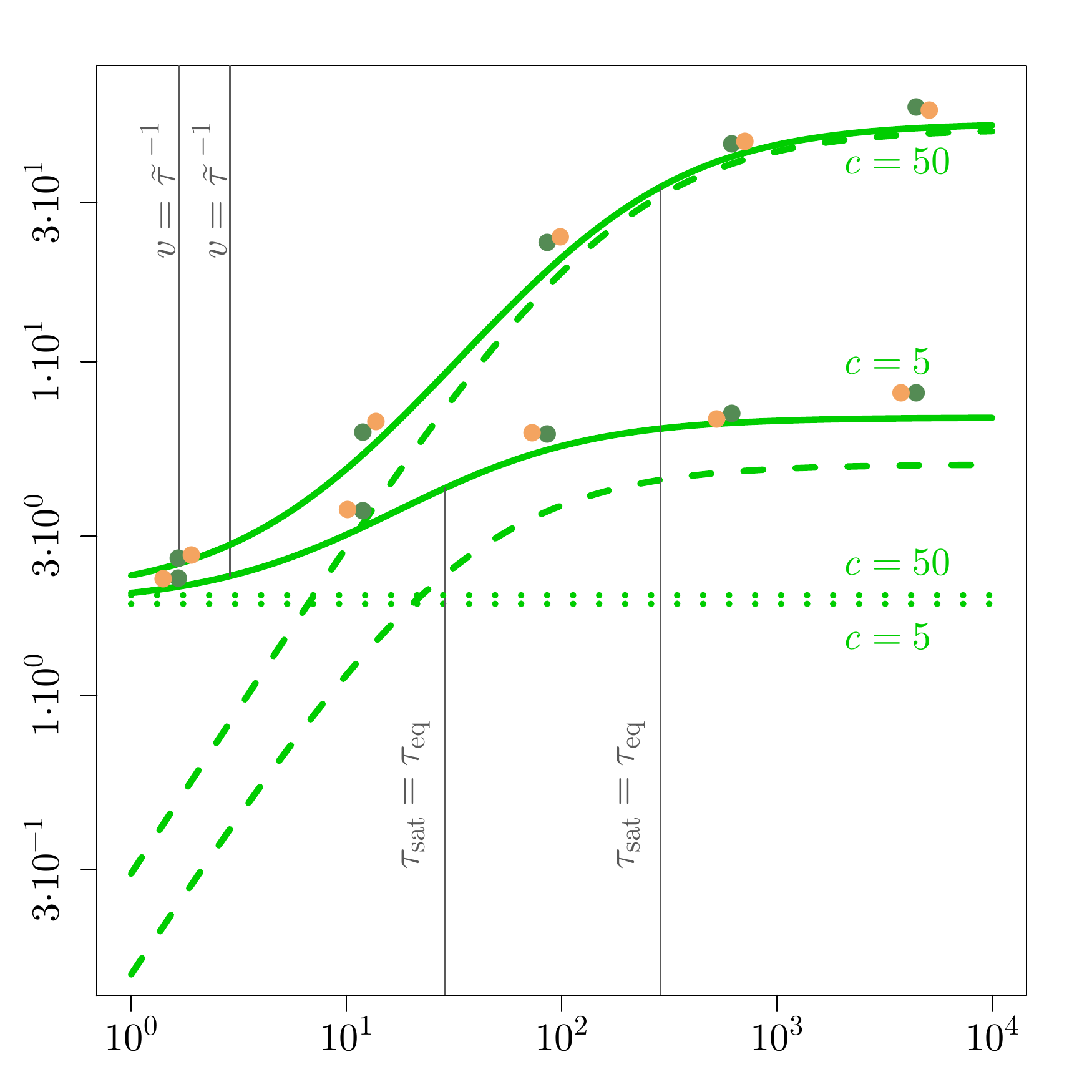} \hspace{1cm}& 
 \begin{sideways} \hspace{40pt} scaled fitness flux, $\ql 2N \phi \qr/\mu$ \end{sideways}& 
\includegraphics[width=.40\textwidth]{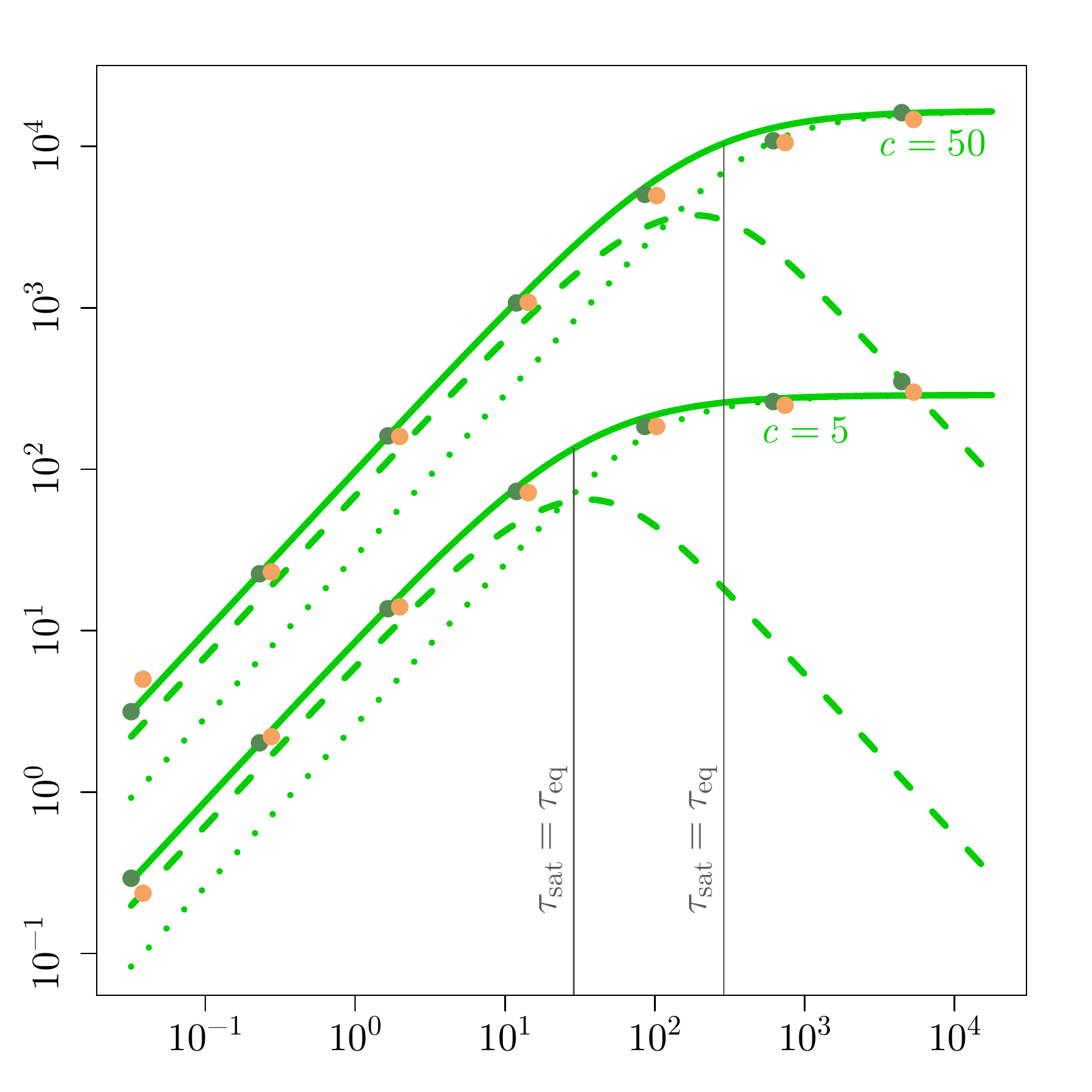}
\\
&  \hspace{50pt} scaled driving rate, $\v /\mu$ & & 
\hspace{50pt} scaled driving rate, $\v/\mu$
\end{tabular}
\caption{{\bf Genetic load and fitness flux.}
(a)~Scaled genetic load $2NL$ (full lines) and its constituents, the adaptive genetic load, $2 N L_\ad$ (dashed lines), and equilibrium genetic load, $2N L_\eq$ (dotted lines), for stationary evolution of non-recombining populations in a fitness seascape. The load components are plotted against the scaled driving rate $\v / \mu$  for stabilizing strengths $c=5, 50$; other parameters like in Fig.~\ref{fig:acfoverDlong}. The analytical results of eq.~(\ref{eq:fullload}) are compared to simulations for diffusive and punctuated fitness seascapes (green and orange dots). The corresponding data for fully recombining populations are shown in Fig.~\ref{fig:freerec}. The genetic load is dominated for $\v \lesssim 1/ \tilde \tau (c)$ by the equilibrium component  and for $\v \gtrsim 1/ \tilde \tau (c)$ by the adaptive component;  it saturates in the micro-evolutionary seascape regime ($\v \gtrsim r^2 / \tau_\eq (c)$). 
(b)~The scaled fitness flux $\langle 2N \phi \rangle$ (solid line) and its components  $\ql 2N\phi_{\rm macro} \qr$ and  $\ql 2N\phi_{\rm micro} \qr$, as defined in eqs.~(\ref{phimacro}), are shown for the same parameters (all flux values are measured in units of $1/\mu$). In macro-evolutionary fitness seascapes, $\langle 2N \phi \rangle$ is an approximately linear function of the driving rate $\v$ and the component $\ql 2N\phi_{\rm macro}\qr$ is the dominant part. In micro-evolutionary seascapes, $\langle 2N \phi \rangle$ saturates and the component $\ql 2N\phi_{\rm micro }\qr$ is the dominant part.
}
\label{loadflux}
\end{figure}

Finally, the different regimes of adaptive trait evolution can be characterized in terms of the genetic load. The adaptive load is asymptotically linear in the driving rate and is subleading to the equilibrium load in the slow-driving regime ($\v \lesssim \tilde \v (c) \equiv1/\tilde \tau (c)$). It becomes dominant for faster driving ($\v \gtrsim \tilde  \v (c)$) and saturates in the  micro-evolutionary regime ($\v \gtrsim r^2 / \tau_\eq(c)$). Fig.~\ref{loadflux}(a) shows this dependence of the adaptive load on the driving rate.

\subsection{Fitness flux} 
\label{sec:Flux}

The fitness flux, $\phi (t)$, characterizes the adaptive response  of a population evolving in a fitness land- or seascape,
\EQ
\phi  (t)  = \int \!  \d E\, f (E,t) \, \partialT \mathcal W(E,t).
\EE
The cumulative fitness flux, $\Phi(\tau) = \int_t^{t+\tau} \phi (t') dt' $, measures the total amount of adaptation over an evolutionary period $\tau$~\cite{Mustonen2007}. 
The evolutionary statistics of this quantity is specified by the fitness flux theorem~\cite{Mustonen2010}.
According to the theorem, the average cumulative fitness flux in a population ensemble measures the deviation of the evolutionary process from equilibrium: this deviation equals the relative entropy of the actual process from a hypothetical time-reversed process~\cite{Mustonen2010, Jarzynski:2006}. It is substantial ---~i.e., the process is predominantly adaptive ---~if $\langle 2 N \Phi \rangle \gtrsim 1$. Specifically, the cumulative  fitness flux of a stationary adaptive process  increases linearly with time, 
$\langle 2N \Phi (\tau) \rangle = \langle \phi \rangle \tau $ with $\langle \phi \rangle > 0$. 

For a quantitative trait in a quadratic fitness seascape of the form (\ref{fE}), we can decompose the fitness flux into contributions of the trait mean and the trait diversity,  
\EQ
\phi  (t) = -2 c_0 \big ( \Gamma (t) - E^* (t) \big ) \frac{\d \Gamma(t)}{\d t} -2 c_0 \frac{\d \Delta(t)}{\d t}.  \label{flux}
\EE
In the stationary population ensemble  (\ref{eq:ansatz1}), the average scaled fitness flux can be expressed in terms of the stationary probability current ${\mathbf J}_\st (\Gamma, E^*)$, 
\EQ
\langle 2N \phi \rangle  =- \frac{2c }{E_0^2}  \int \! \d \Gamma \d E^* \, (\Gamma - E^*) J_\st ^\Gamma (\Gamma, E^*),   \label{fluxflux}
\EE 
where $J_\st ^\Gamma (\Gamma,E^*)$ is the $\Gamma$--component of ${\mathbf J}_\st(\Gamma,E^*)$.
The fitness flux can be computed analytically from eq.~(\ref{J}),
\EQ
\langle  2N\phi \rangle (c,\v, r^2)  =  2 c \v \,\xv .
\EE
In the regime of substantial stabilizing strength ($c\gtrsim1$), we get

\EQ
\langle  2N\phi \rangle (c,\v, r^2) \simeq 
\left \{
\begin{array}{cc}
2c \v  \Big[1 - \O \Big (\dfrac{\tau_\eq}{\tau_\sat} \Big )\Big] &  \macro,
\\ \\
\dfrac{4  c^2 r^2}{\tilde \tau(c)} \, \Big [1 - \O\Big (\dfrac{\tau_\sat}{\tau_\eq} \Big ) \Big]   &  \micro,
\end{array} \right. 
\label{eq:flux_macro}
\EE
where the drift time $\tilde \tau (c)$ is given by eqs.~(\ref{tautilde}) and~(\ref{deltaconstraint}). The fitness flux depends linearly on the driving rate in a macro-evolutionary fitness seascape, and it saturates in the regime of micro-evolutionary fitness fluctuations. Fig.~\ref{loadflux}(b) shows this dependence of the fitness flux on the driving rate.

We can express the fitness flux in terms of correlation functions of the trait mean $\Gamma(t)$ and the lag $\Lambda(t)$, 
which results in a simple relation between fitness flux and adaptive load. Inserting  the probability current of eq.~(\ref{def:J}) into the integral of eq.~(\ref{fluxflux}), we find
\EQA
\ql 2N\phi\qr &=&
\frac{2c^2 \ql \delta \qr}{E_0^2} \Big( \ql \Lambda^2 \qr - \ql \Lambda^2\qr_\eq \Big )   
+ \frac{4c \theta}{E_0^2} \lim_{\tau \searrow 0} 
\Big( \ql \Lambda(t+\tau) (\Gamma(t)-\Gamma_0) \qr -  
\ql \Lambda(t+\tau) (\Gamma(t)-\Gamma_0) \qr_\eq  \Big ) 
\nonumber \\
& = & c\ql \delta \qr \ql2 N   L\qr_{\rm ad}(c,v,r^2) [1 + \O(\theta)]. 
\label{philambda} 
\EEA 
From this representation, we obtain the spectral decomposition of the fitness flux,
\EQ
\langle  2N\phi \rangle (c,\v, r^2) 
 =   \int_0^\infty \langle 2N \phi (\omega) \rangle \, d\omega
\label{phi_spectral}
\EE
with 
\EQA
\langle 2N \phi (\omega) \rangle  &=&  2c \v\frac{c \ql \delta \qr  }{{\pi/2}}\frac{\omega^2}{(\tau_\eq^{-2}(c) + \omega^2)(\tau_{\rm sat}^{-2}(v,r^2)+ \omega^2)}[1+\O(\theta/(c\ql \delta \qr)].
\EEA
Using a cutoff frequency $\omega_c = k / \tau_\eq (c)$ with a constant $k$ of order 1, we can now define a macro-evolutionary flux component,
\EQA
\langle  2N\phi_{\rm macro} \rangle &=&  \int_0^{\omega_c} \langle 2N \phi (\omega) \rangle \, d\omega \nonumber \\
& = & 2c \v \; \xv \frac{ ({\tau_\eq^{-1}(c)+2\mu})\arctan [k ] -  ({\tau_{\rm sat}^{-1}(\v,r^2)+2\mu}) \arctan \left[k \, \tTwo/\tOne \right] }{(\pi/2)(\tau_\eq^{-1}(c)-\tau_{\rm sat}^{-1}(\v,r^2))},  \nonumber \\
\label{phimacro}
\EEA
and the complementary micro-evolutionary component 
\EQA
 \langle  2N\phi_{\rm micro} \rangle =  \int_{\omega_c}^\infty \langle 2N \phi (\omega) \rangle \, d\omega = \langle  2N\phi \rangle-\langle  2N\phi_{\rm macro} \rangle.
\EEA
 In  the regime of substantial stabilizing selection ($c\gtrsim1$), the macro-evolutionary fitness flux in (\ref{phimacro}) reads
\EQA
\langle  2N\phi_{\rm macro} \rangle (c,\v, r^2) &\simeq&\begin{cases} 2c \v \,\dfrac{2}{\pi} \arctan [k]   ,   &  \macro \\ \\
2 c \v \, \dfrac{\tau_\sat^2 (\v,r^2)}{\tau_\eq^2 (c)}  \,  \dfrac{2}{\pi} \big(k-  \arctan [k]\big) \sim \dfrac{1}{v} &  \micro.
\end{cases} \label{phimacro_regimes}
\EEA
This fitness flux component quantifies the macro-evolutionary part of adaptation. In macro-evolutionary fitness seascapes ($\tau_\sat (\v,r^2) \gtrsim \tau_\eq(c)$), it increases proportionally to the driving rate~$\v$ and, for $k > 1$, it  represents the main fraction of the total fitness flux $\ql 2N\phi\qr$. In micro-evolutionary fitness seascapes ($\tau_\sat (\v,r^2) \lesssim \tau_\eq(c)$), this component is suppressed: the macro-evolutionary  fitness flux does not carry  information on rapid fitness fluctuations. This cross-over of $\langle  2N\phi_{\rm macro} \rangle$ and of the complementary component $\langle  2N\phi_{\rm micro} \rangle$ is shown in Fig.~\ref{loadflux}(b). 

The spectral decomposition of the fitness flux has  important consequences for the analysis of macro-evolutionary adaptation. The detection of a substantial cumulative  fitness flux $\langle  2N\Phi_{\rm macro} (\tau) \rangle > 1$ over a macro-evolutionary period~$\tau$ is not confounded by the simultaneous presence of  micro-evolutionary (for example seasonal) fitness fluctuations. Since the cumulative fitness flux is a measure of  entropy production  during adaptation, the spectral decomposition~(\ref{phi_spectral}) also has an important information-theoretic interpretation: The difference $\langle  2N\phi_{\rm micro} \rangle = \langle  2N\phi \rangle - \langle  2N\phi_{\rm macro} \rangle$ is the average loss of information per unit time through temporal coarse-graining. This loss is a non-equilibrium analogue of the entropy production  by spatial coarse-graining.

\subsection{Predictability and entropy production}
In ref.~\cite{Nourmohammad2012}, we quantified the evolutionary predictability of the molecular traits  across an ensemble of populations by
\EQ
\mathcal P \equiv \exp \big [\langle{ S(\W)} \rangle - S( \langle W \rangle) \big ]
\label{calP},
\EE
with $S (\W)  \equiv - \int \W(E) \, \log \W(E) dE$.
This definition compares the ensemble-averaged ``micro-evolutionary'' Shannon entropy of the phenotype distribution within a population, 
$
\langle S{(\W)} \rangle \equiv \int_{\W}  S(\W) \, Q(\W), 
$
and the ``macro-evolutionary'' Shannon entropy of the mixed distribution,
$
S(\langle \W \rangle ) \equiv S \big ( \int_{\W}  \W \, Q(\W) \big ),
$
which is obtained by compounding the trait values of all populations into a single distribution.
 We have shown that the predictability is generically low in a neutral ensemble, but stabilizing selection in a single fitness landscape can generate an evolutionary equilibrium wiht predictability values $\mathcal P$ of order 1~\cite{Nourmohammad2012}. 
 
Here we compute the predictability in a time-dependent ensemble of populations that descend from a common ancestor population. Similarly to ref.~\cite{Nourmohammad2012}, we evaluate eq.~(\ref{calP}) for a distribution $Q_t(\W)$ with the initial condition $Q_{t_a} (\W) = \delta(\W - \W_a)$ at time $t_a = t - \tau/2$. We obtain the time-dependent predictability

\EQ
\mathcal P (\tau; c, \v, r^2) \simeq \left(\frac{\langle \delta \rangle (c) }{\langle d ^{(2)}\rangle (\tau; c, \v, r^2)/2 + \langle \delta \rangle (c) }\right)^{1/2}= \left(\frac{1}{1+\Omega^{(2)} (\tau; c, \v, r^2) / 4 \theta}\right)^{1/2}.   
\label{calP_eq}
\EE
Here, $\Omega^{(2)} (\tau; c,v,r^2) \equiv 2 \theta \, \langle d^{(2)} \rangle (\tau; c, \v, r^2) /  \langle \delta \rangle  $ denotes the  ratio between trait divergence  and diversity for the descendent populations.  The trait statistics in a macro-evolutionary fitness seascape, given by eqs.~(\ref{dkappa}) and~(\ref{deltaconstraint}), entail the evolutionary  predictability 
\EQA
\mathcal P (\tau; c, \v, r^2) &=&\mathcal P_\eq (c) \left [ 1 - \frac{1}{2}  \v \, \tilde \tau \frac{\tau -  2\tau_\eq (c)}{2N}  \,   \Big[ 1+ \O\left(\tau \,\tau_\eq(c)c v/N , \tau/ \tTwoA,\theta/(c \ql \delta \qr)\right) \Big] \right] 
\EEA  
 for $\tau \gtrsim \tau_\eq (c)$, with $\mathcal P_\eq (c) = (1+\x/(2c))^{-1/2}=(1+1/(2c))^{-1/2}[1+\O(\theta/(c\ql\delta\qr))]$. There are two stochastic components   that generate  macro-evolutionary entropy and, hence,  reduce the evolutionary predictability: fluctuations induced by genetic drift on short time-scales $\tau \lesssim \tau_\eq (c)$ and fluctuations of the fitness peak over time-scales $\tau \gtrsim \tau_\eq (c)$. Stabilizing selection, on the other hand, reduces the entropy production of the adaptive process~\cite{Nourmohammad2013}. Therefore, the predictability of an adaptive process in a fitness seascape with a substantial stabilizing strength and sufficiently slow driving rate can remain of order one.

Parallel and convergent evolution at the functional level, paired with strongly divergent genome evolution has been observed in a number of recent experiments~\cite{Tenaillon2012, Toprak:2012ff, Barroso13}. These experimental observations can be explained in a natural way, if we assume that many of these functions involve a complex quantitative trait.

\section{Inference of adaptive trait evolution}
\label{sec:Inference}

The statistical theory developed in this paper suggests a new method to infer selection on quantitative traits. Our method is based on  trait evolution in a single-peak fitness seascape, as defined in eq.~(\ref{fE}), which is parametrized by its stabilizing strength $c$ and its driving rate~$\v$. 

Two main results are relevant for the inference of selection. First, evolution in a macro-evolutionary fitness seascape affects the population mean trait in complementary ways: it generates conservation on shorter scales and adaptation on longer scales of evolutionary time. These characteristics are measured by the expected trait divergence between populations,~$\langle D^{(\kappa)}(\tau) \rangle$, which depends on the divergence time $\tau$ and on the selection parameters $c$ and $\v$ in a characteristic way. The divergence can be measured either between an ancestral population and a descendent population ($\kappa = 1$) or between two descendent populations evolving from a common ancestor population ($\kappa = 2$). As discussed in section~3.2, these measures are generically distinct for adaptive processes\footnote{The relative difference between $\langle D^{(1)}(\tau) \rangle$ and $\langle D^{(2)}(\tau) \rangle$ is small (Fig.~5). This difference is conceptually important, however, because it manifests the violation of detailed balance in adaptive processes. Similar effects are ubiquitous in divergence data of trait adaptation across multi-branch phylogenies.}.
Second, the expected trait diversity within populations,~$\langle \Delta \rangle$, shows a weaker signal of conservation.  Moreover, it decouples from the adaptive process in a single-peak fitness seascape over a wide range of evolutionary parameters, as discussed in section~\ref{chap:delta}. 

\begin{figure}[t] 
\centering
\boxed{
\includegraphics[width=0.5\textwidth]{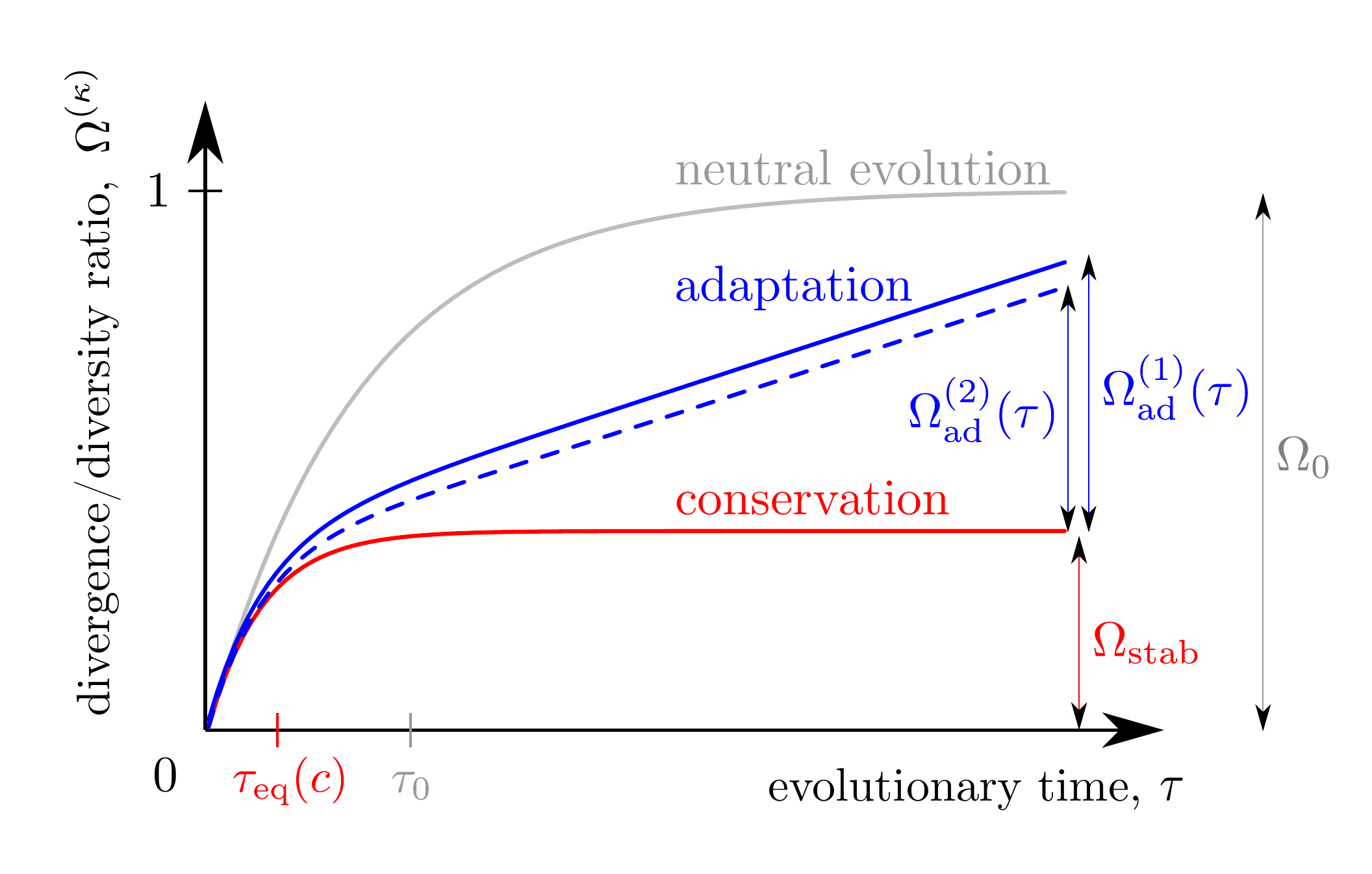} 
}
\caption{ {\bf The universal divergence-diversity ratio $\Omega^{(\kappa)}$} ($\kappa = 1,2$), as defined in eq.~(\ref{defOmega}), for a quantitative trait evolving in a single-peak fitness land- or seascape. This ratio is plotted as a function of the scaled divergence time, $ \tau$. 
{\em Neutral evolution}: The function $\Omega_0 (\tau)$ is independent of $\kappa$ and reaches its saturation value 1 on times scales $ \tau \gg \tau_0 = 1/\mu$ (grey curve). 
{\em Conservation in a fitness landscape}: The function $\Omega_\eq (\tau)$ is independent of $\kappa$ and has a smaller saturation value $\Omega_{\rm stab} (c)$ reached faster than for neutral evolution, on time scales $\tau \gg \tau_\eq (c)$ (red curve). 
{\em Adaptation in a fitness seascape}: There is a linear surplus $\Omega_\ad^{(\kappa)} (\tau) \simeq \v [\tau - \kappa \tau_\eq (c)]$, which measures the amount of adaptation (blue curves). 
}
\label{fig:inference2}
\end{figure}

Our test statistics is the time-dependent divergence-diversity ratio
\EQ
\Omega^{(\kappa)} (\tau)  = 2 \theta \frac{\langle D^{(\kappa)}(\tau) \rangle}{\langle \Delta \rangle}
 \qquad (\kappa=1,2),  
\label{defOmega}
\EE
where $\theta = \mu N$ denotes the nucleotide diversity.  This function depends on the divergence time $\tau$ and on the selection parameters $c$ and $\v$. The typical behavior of $\Omega^{(\kappa)} (\tau)$ for   different evolutionary modes is shown in Fig.~\ref{fig:inference2} and can be summarized as follows:
\begin{itemize}
\item {\em Neutral evolution} ($c = 0$). The divergence-diversity ratio has an initially linear increase due to genetic drift and approaches a maximum value 1 with a relaxation time $\tau_0 = 1/\mu$, 
\EQA
\Omega^{(\kappa)} (\tau) & = & \Omega_0 (\tau) \simeq
\left \{
\begin{array}{ll}
\mu \tau   & \mbox{ for $\tau \ll \tau_0$} 
\\
1 & \mbox{ for $\tau \gg \tau_0$}
\end{array} \right.  \qquad (\kappa=1,2).
\label{Omega_neutral}
\EEA
The function $ \Omega_0 (\tau)$, which does not depend on $\kappa$ by detailed balance, is shown as a grey line in Fig.~\ref{fig:inference2}. 

\item{\em Conservation in a fitness landscape} ($c \gtrsim 1, \v = 0$).
The divergence-diversity ratio approaches a smaller maximum value, $\Omega_{\rm stab} (c) < 1$, with a proportionally shorter relaxation time $\tau_\eq (c) = \Omega_{\rm stab} (c) / \mu$, 
\EQA
\Omega^{(\kappa)} (\tau) & = & \Omega_\eq (\tau;  c) \simeq
\left \{
\begin{array}{ll}
\mu \tau  & \mbox{ for $\tau \ll \tau_\eq (c)$} 
\\
\Omega_{\rm stab} (c) & \mbox{ for $\tau \gg \tau_\eq (c)$}
\end{array} \right.   \qquad (\kappa=1,2).
\label{Omega_cons}
\EEA
The function $ \Omega_\eq (\tau;  c)$, which does not depend on $\kappa$ by detailed balance, is shown as a red line in Fig.~\ref{fig:inference2}. Over a wide range of evolutionary parameters, the maximum value depends on the stabilizing strength in a simple way, $\Omega_{\rm stab} (c) \sim 1/(2c)$, with corrections for weaker selection and for larger nucleotide diversity. 

\item{\em Adaptation in a macro-evolutionary fitness seascape} ($c \gtrsim 1, 0 < \v \lesssim 1/\tilde \tau $).
The divergence-diversity ratio acquires an adaptive component, 
\EQA
\Omega^{(\kappa)} (\tau) & = & \Omega_\eq (\tau; c) + \Omega_\ad^{(\kappa)} (\tau; \v) 
\nonumber \\
& = & \Omega_\eq (\tau; c)  + \frac{ \v}{2} \,[\tau-\kappa\tau_\eq(c)]
\qquad (\kappa=1,2) ,   
\label{Omega_ad}
\EEA
with corrections for weaker selection and for $\tau$ approaching the non-equilibrium saturation time $\tau_{\rm sat} = r^2/\v$. The functions $\Omega^{(\kappa)} (\tau)$ are shown as blue lines in Fig.~\ref{fig:inference2}. 
\end{itemize}

The $\Omega$-test for selection on quantitative traits is conceptually related to the McDonald-Kreitman test for adaptive sequence evolution~\cite{Kreitman:1991vh}; both tests are based on a comparison between divergence and diversity. However, the $\Omega$ statistics does not require a corresponding ``null trait'' that evolves near neutrality and takes the role of synonymous sequences. Indeed, no such neutral trait gauge is available in most cases. The~$\Omega$-test instead evaluates {\em time-resolved} divergence data $D(\tau)$. In macro-evolutionary fitness seascapes, this test infers stabilizing and directional selection using their different characteristic time scales.  
In principle, a single data point in the saturation regime $\tau \gtrsim \tau_\eq (c)$ determines the stabilizing strength of a fitness landscape, and two data points are sufficient to determine strength and driving rate of a fitness seascape. Our statistical theory also specifies an error model for the probabilistic inference of seascape parameters from noisy data, based on the statistics of the finite time propagator detailed in Appendix A. This inference method can be generalized from single lineages to multi-species phylogenies and will be discussed in a forthcoming empirical study. 

An important prerequisite for the wide applicability of the $\Omega$ test is its {\em universality}: the divergence-diversity ratio depends on the selection parameters $c$ and $\v$, but it decouples from the trait's genetic basis. In particular, it depends only weakly on the number and trait amplitudes of the constitutive sequence sites, and on the amount of recombination between these sites. All of these genetic factors are, in general, unknown. They act as confounding factors on non-universal observables such as the trait divergence and diversity, which have been used in previous studies to infer selection~\cite{Gilad:2006kt, Bedford:2009fy}. The $\Omega$ statistics also decouples from details of the selection dynamics; it can be applied to continual as well as to punctuated adaptive processes. We have tested this universality by extensive numerical simulations, which are reported in Appendix~B.

\section{Discussion} 

In this paper, we have developed a statistical theory for the evolution of a quantitative trait in a stochastic fitness seascape. 
The fitness model used for our analysis, a single-peak seascape with diffusive or punctuated peak displacements, covers a broad spectrum of biologically relevant evolutionary scenarios~\cite{Nourmohammad2013}. The two seascape parameters $c$ and $\v$ quantify stabilizing and directional selection on the trait, which, in turn, govern the trait's fundamental evolutionary modes of conservation and adaptation. Our analysis shows that these modes are not mutually exclusive, but are joint features of dynamic selection models. 

In a macro-evolutionary fitness seascape, conservation and adaptation are associated with different time scales: conservation is observed on shorter scales, while adaptive changes build up on longer scales of evolutionary time. Micro-evolutionary fitness fluctuations, on the other hand, lead to reduced genetic adaptation, which decouples from the macro-evolutionary dynamics of the trait. Rapid adaptive response to seasonal or other fluctuations of the environment often involves epigenetic modifications or phenotypic switching\cite{Rivoire2013}. The evolutionary roles of these mechanisms are beyond the scope of this paper. The spectral decomposition of the fitness flux, which has been introduced above, quantifies how the adaptive process is distributed on different scales of evolutionary time. 

Our theory suggests a new test for selection on quantitative traits, which has important potential applications. At the sequence level, an increasingly complex picture of selection has emerged in recent years. Notably, we have acquired a growing repertoire of empirical genotype-fitness landscapes~\cite{Szendro13}, which has generated important experimental and theoretical insights into the evolutionary dynamics on these landscapes. However, we still know little about the statistical properties of empirical phenotype-fitness maps, and next to nothing about phenotype-dependent seascapes. Systematic inference of selection on molecular quantitative traits, such as levels of gene expression and enzymatic activity, can contribute to close this gap. Eventually, fitness land- and seascapes for individual traits will need to be integrated into larger phenotype-fitness maps, which include fitness interactions between traits.

\cleardoublepage

\setcounter{table}{0}
\renewcommand{\thetable}{A.\arabic{table}}
\setcounter{figure}{0}
\renewcommand{\thefigure}{A.\arabic{figure}}

\begin{appendix}

\section{Analytical theory of the adaptive ensemble}
\label{chap:solveG}
In Section \ref{chap:gamma}, we obtained the Gaussian stationary distribution $Q_\st(\Gamma,E^*)$ in a diffusive seascape from the underlying Fokker-Planck equation~(\ref{diffGamma}). Here we use a Langevin representation to compute the time-resolved trait divergence $\ql d^{(\kappa)} \qr(\tau)$  ($\kappa = 1,2$). This derivation, which reproduces mean and variance of the distribution $Q_\st(\Gamma,E^*)$, applies to diffusive and punctuated fitness seascapes. We also compute the full propagator function $G_\tau(\Gamma, E^*|\Gamma_a, E^*_a)$ for macro-evolutionary diffusive seascapes. The propagator in a punctuated fitness seascape has the same mean and variance, but differs in higher trait moments. 

\paragraph{Moments of the optimal trait.}
In a  diffusive seascape, the fitness peak $E^*(t)$ follows  an Ornstein-Uhlenbeck process with Langevin representation
 \EQA
\partial_t E^*(t) =   -\frac{\v}{r^2} (E^*(t) - \cE) + \eta(t) ,
\label{eq:langEstar}
\EEA
where $\eta(t)$ is a  Gaussian random variable with the statistics
\EQ
\ql \eta (t) \qr = 0, \quad \ql \eta(t) \eta(t') \qr = 2 \v E_0^2 \, \delta(t-t') . 
\label{eq:eta_mom}
\EE
Formally solving eq.~(\ref{eq:langEstar}), 
\EQ
E^*(t+\tau)= E^*(t)e^{- \tau/\tTwoA}+ \mathcal E (1-e^{- \tau/\tTwoA}) + \int_{t_1}^{t_2} \!\text d t'\, e^{- (\tau-t')/\tTwoA} \eta(t'),
\label{EStarDiff}
\EE
and evaluating the noise correlations (\ref{eq:eta_mom}), we obtain the average peak value with an initial condition $E^*(t) = E_a$ and the autocorrelation function of the fitness peak in the stationary ensemble, 
\EQA
\langle E^*(t+\tau) \rangle (E_a) & = & E_a e^{- \tau/\tTwoA} + \mathcal E\big(1-e^{-\tau/\tTwoA}\big) ,
\label{prop:mom1}\\
\ql E^*(t)E^*(t+\tau) \qr &=& \mathcal E^2 + E_0^2 r^2 e^{-\tau/\tTwoA}.  
\label{EStarCor}
\EEA
It is straightforward to check that eqs.~(\ref{prop:mom1}) and (\ref{EStarCor}) are valid also for punctuated seascapes. 

\paragraph{Moments of the trait mean.} 
The Langevin equation for $\Gamma (t)$ reads 
\EQA
\partial_t \Gamma(t) = -2\mu(\Gamma(t)-\Gamma_0)-\ql\Delta \qr\frac{2c}{E_0^2}(\Gamma(t)-E^*(t))+\xi(t), \label{eq:langE}
\EEA
where $\xi(t)$ is a Gaussian noise with  the statistics 
\EQ
\ql \xi(t)\qr=0,\quad \ql \xi(t) \xi(t')\qr= \frac{\ql \Delta \qr}{N} \, \delta(t-t'),
\quad \ql \xi(t) E^*(t')\qr=0. 
\label{eq:xi_mom}
\EE 
For diffusive seascapes, the last term in~\ref{eq:xi_mom} is equivalent to $\ql \xi(t) \eta(t')\qr=0$, which implies that genetic drift and fitness seascape fluctuations are independent. The formal solution of eq.~(\ref{eq:langE}) reads
\EQA
\nonumber\Gamma(t+\tau) &=& e^{-\tau/\tOneA} \Gamma(t) + (1\!-\!\x)(1- e^{-\tau/\tOneA}) \Gamma_0\\
&&+\int_{t}^{t+\tau} \text d t' \,  ( E^*(t') c \ql \delta \qr  + \xi(t'))\,e^{-( t+\tau-t')/\tOneA}, 
 \label{Gamma_sol}
\EEA
where $\x=[1+2\theta/(c \ql \delta \qr)]^{-1}$. In the case of a diffusive fitness seascape, we can insert the trajectory of the fitness peak $E^*(t)$ given by  eq.~(\ref{EStarDiff}),
\EQA
\nonumber\Gamma(t+\tau) &=&\Gamma(t) e^{-\tau/\tOneA}  +  E^*(t)\xmv\big(e^{-\tau/\tTwoA}-e^{-\tau/\tOneA}\big) + \Gamma_0(1\!-\!\x)(1- e^{-\tau/\tOneA}) \\
&&+\mathcal E \xmv \left[\big( 1- e^{-\tau/\tTwoA} \big )+ \frac{\tOneA}{\tTwoA}\big( 1-e^{-\tau/\tOneA}\big) \right ] \label{GammaTauNoise}
\\&& +\int_{t}^{t+\tau}\! \d t' \,\left[\xi(t') e^{-(t+\tau-t')/\tOneA} + \eta(t') \xmv \big( e^{- (t+\tau-t')/\tTwoA}-e^{- (t+\tau-t')/\tOneA} \big)  \right]; \nonumber
\EEA 
see, e.g., section~4 of \cite{Gardiner:2004tx}. Evaluating the noise correlations (\ref{eq:xi_mom}), we obtain
\EQA
\ql \Gamma \qr & = & w(c) \mathcal{E} + (1 - w(c)) \Gamma_0,
\label{Gamma1}
\\ \nonumber \\
\ql \Gamma(t) \Gamma(t+\tau) \qr &=& \ql \Gamma \qr^2+ \ql \hat \Gamma^2 \qr  e^{-\tau/\tOneA} + r^2 E_0^2 \xv\xmv (e^{-\tau/\tTwoA} - e^{-\tau/\tOneA}),  \label{GammaCor}
 \\ \nonumber \\
 \ql \Gamma(t+\tau)E^*(t) \qr &= &\ql \Gamma \qr \mathcal E + E_0^2  r^2 \xmv e^{-\tau/\tTwoA} \nonumber \\
&&- \theta(\tau) E_0^2 \, \v \tau_\eq (c) \xv\xmv \big(e^{-\tau/\tOneA}-e^{-\tau/\tTwoA}\big),
\label{GammaEstarCor}
\EEA
where $\xv=[1+(2\theta+2N\tau_\st^{-1}(v,r^2))/(c\ql\delta\qr)]^{-1}$ and $\theta(\tau)$ is the unit step function; i.e., $\theta (\tau)  = 1$ for $\tau > 0$ and $\theta (\tau) = 0$ otherwise. The relations (\ref{Gamma1}) -- (\ref{GammaEstarCor})  are also valid for punctuated seascapes, as can be shown by evaluating eq.~(\ref{Gamma_sol}) with the noise terms~(\ref{prop:mom1}),~(\ref{EStarCor}), and~(\ref{eq:xi_mom}). The time-reflection asymmetry of the cross-correlation (\ref{GammaEstarCor}) reflects the causal relation between $\Gamma$ and $E^*$. The equal-time correlations reproduce the moments (\ref{eq:Sigma.full}) obtained from the solution of the Fokker-Planck equation. 

From the autocorrelation function (\ref{GammaCor}), we immediately obtain the scaled divergence $\langle d^{(1)} \rangle$  reported in eq.~(\ref{dkappa}). For the divergence between descendent populations, $\langle d^{(2)} \rangle$, we additionally use the fact that the fitness fluctuations  in the different lineages are independent  of each other. In a diffusive fitness seascape, we have 
\EQ
 \langle \eta_i(t)\eta_j(t')\rangle =\delta_{i,j}\, \delta (t-t')\, 2 \v E_0^2, \qquad i,j=1,2, 
\EE
which implies
\EQ
\big\langle( E_1^*(t+\tau_1)-\langle E_1^*(t+\tau_1)\rangle )( E_2^*(t+\tau_2)-\langle E_2^*(t+\tau_2)\rangle ) \big\rangle 
= 0;
\EE
the latter relation is valid also for punctuated seascapes. 

\paragraph{Propagators.}
We recall the decomposition of the bivariate propagator,
\EQA
G_\tau(\Gamma, E^* |\Gamma_a, E^*_a)= G_\tau(\Gamma | \Gamma_a, E^*_a,  E^*) \,G_\tau(E^*|E^*_a) \label{TraitProp}, 
\EEA
which reflects the independence of the fitness peak dynamics from the trait mean. 

The fitness peak propagator takes the standard form for an Ornstein-Uhlenbeck process and a Poisson jump process, respectively, 
\EQA
G_\tau(E^*|E^*_a) = \left \{
\begin{array}{ll}
\dfrac{1}{\sqrt{2\pi \ql \hat E^{*2}\qr ({\tau, E_a^*}) }}\exp\left[ -\dfrac{(E^*-\ql E^*\qr(\tau, E_a^*) )^2}{2 \ql \hat E^{*2}\qr (\tau) }\right] &  \text{(diffusive seascape)}
\\ \\
e^{-\tau/\tTwoA}\delta(E^*\!-\!E^*_a)+(1\!-\!e^{-\tau/\tTwoA})R_\eq(E^*) & \text{(punctuated seascape)} ;
\end{array} \right. 
\label{prop:punct}
\EEA
see, e.g., ref.~\cite{Gardiner:2004tx}. In both cases, the propagator has the same mean and variance, 
\EQ
\ql E^*\qr(\tau, E_a^*) = E_a e^{- \tau/\tTwoA} + \mathcal E\big(1-e^{-\tau/\tTwoA}\big) ,
\qquad 
\langle \hat E^{* 2} \rangle (\tau)
= r^2 E_0^2 \big(1-e^{-\tau/\tTwoA}\big) ,
\EE
in accordance with eqs.~(\ref{prop:mom1}) and (\ref{EStarCor}). 

For diffusive seascapes, we can also compute the Gaussian propagator of the trait mean for given fitness peak positions, 
\EQA
G_\tau(\Gamma|\Gamma_a,E_a^*,E^*) = \frac{1}{\sqrt{2\pi \ql \hat \Gamma^2 \qr ( \tau)}} \exp\left[-\frac{ 1}{ 2} \frac{\big(\Gamma-\ql \Gamma \qr(\Gamma_a,E_a^*,E^*,\tau)\big)^2}{\ql \hat \Gamma^2 \qr (\tau)} \right]. \label{Gamma2Gamma1E1E2}
\EEA
If $\tau_\eq \lesssim  \tau_{\text{sat}}(\v,r^2)$ and $\tau \lesssim   \tau_{\text{sat}}(\v,r^2)$, we can approximate the  stochastic trajectory of the trait optimum $E^*(t')$ in the time interval $t_a = t - \tau \leq t' \leq t$ by the most likely trajectory for given initial and the final values: $E^*(t')= E_a^* + ((t' - t_a)/\tau) (E^*-E^*_a)$. In this saddle-point approximation, we obtain the conditional trait moments
 \EQA
 \langle \Gamma\rangle(\Gamma_a,E_a^*,E^*, \tau)& = &\Gamma_a \e^{-\tau/\tOneA} +\big(E^*_a  \x +\Gamma_0 (1-\x) \big)    \big (1-\e^{-\tau / \tOneA} \big )  
\nonumber \\
&&+ \frac{E^*-E^*_a}{\tau}\, \x \, \big [{\tau}{}-\tOneA\big(1-\e^{-\tau /\tOneA}\big)\big],\\
\nonumber && \\
\langle \hat \Gamma^2\rangle (\tau)  &=&E_0^2 \frac{\x}{2c} \big(1-\e^{-\tau /\tOneA}\big).
\label{Gamma_cond}
\EEA
Equations (\ref{TraitProp}) -- (\ref{Gamma_cond}) determine the joint propagator $G_\tau(\Gamma, E^*|\Gamma_a, E^*_a)$ for divergence times $\tau_\eq \lesssim  \tau_{\text{sat}}(\v,r^2)$. In the large-time limit, $\tau \gg \tau_\sat (\v,r^2)$, the propagator becomes independent of the initial condition and approaches  the stationary distribution, $G_\tau(\Gamma,E^*|\Gamma_a, E^*_a) \simeq Q_\st(\Gamma,E^*)$, given by eqs.~(\ref{eq:ansatz1}--\ref{eq:Sigma.full}). In most biological experiments,  the trait optimum values are hidden variables of the evolutionary process. In that case, the only observable propagator is the marginal propagator for the trait mean, 
\EQA
G_\tau(\Gamma | \Gamma_a) & \equiv & \int d E_a^* d E^*\,G_\tau(\Gamma, E^*|\Gamma_a, E^*_a)
\, \frac{Q_\st (\Gamma, E_a^*)}{Q_\st (\Gamma)}
\nonumber \\
& = &  \frac{1}{\sqrt{2\pi \langle D^{(1)}\rangle  ( \tau)}} \exp\left[-\frac{ 1}{ 2} \frac{(\Gamma - \langle \Gamma \rangle (\Gamma_a, \mathcal{E}, \tau) )^2}{ \langle D^{(1)} (\tau) \rangle} \right] .
\EEA

\section{Numerical simulations}

We test our analytical results by simulations of a Fisher-Wright process  for the evolution under neutral mutation-drift dynamics, in fitness landscapes with  stabilizing selection, and  in  diffusive or in punctuated fitness seascapes  for sexual and asexual populations. We evolve a population of $N$ individuals with genomes $\a^{(1)}, \dots, \a^{(N)}$, which are bi-allelic sequences of length $\ell$. A genotype $\a$ defines a phenotype $E(\a) = \sum_{i=1}^\ell E_i a_i$; the phenotypic effects $E_i$ are drawn from various distributions (see below). In each generation, the sequences undergo point mutations with a probability $\epsilon \mu$ per generation, where $\epsilon$ is the generation time. The sequences of next generation are then obtained by multinomial sampling; the general form of the sampling probability  is proportional to $[1 + \epsilon f(E(\a),t)]$, with the fitness seascape $f(E, t)$ given by (\ref{fE}). 
For a diffusive seascape, a new optimal trait value $E^*(t)$ is drawn before each reproduction step from a Gaussian distribution with mean $(1-\epsilon \v/r^2)E^*(t)+\epsilon (\v/r^2)\mathcal E$ and variance $\epsilon \v E_0^2$.  For a punctuated seascape, a new, uncorrelated fitness peak is drawn from the distribution $R_\eq(E^*)$ with probability $\epsilon \v/r^2$.

\begin{figure}[t]
\centering 

\begin{tabular}{rlrl}
 &\fig{a}&& \fig{b}  \\
\begin{sideways} \hspace{70pt} 
   divergence, $\ql d ^{(1)}\qr(\tau) $\end{sideways} &
\includegraphics[width=0.40\textwidth]{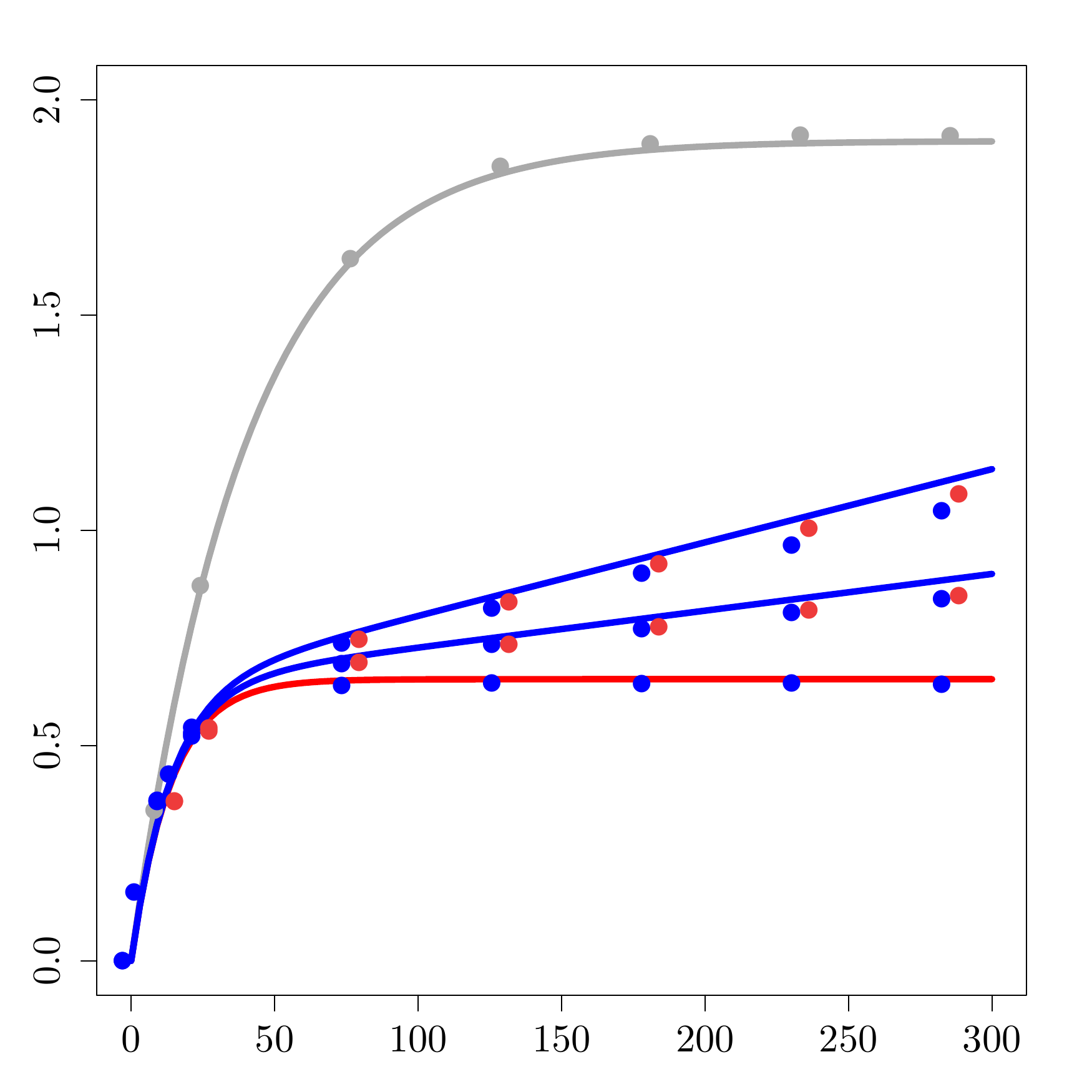} & 
\begin{sideways} \hspace{70pt} 
   divergence, $ \ql d^{(1)}\qr(\tau) $\end{sideways} &
\includegraphics[width=.40\textwidth]{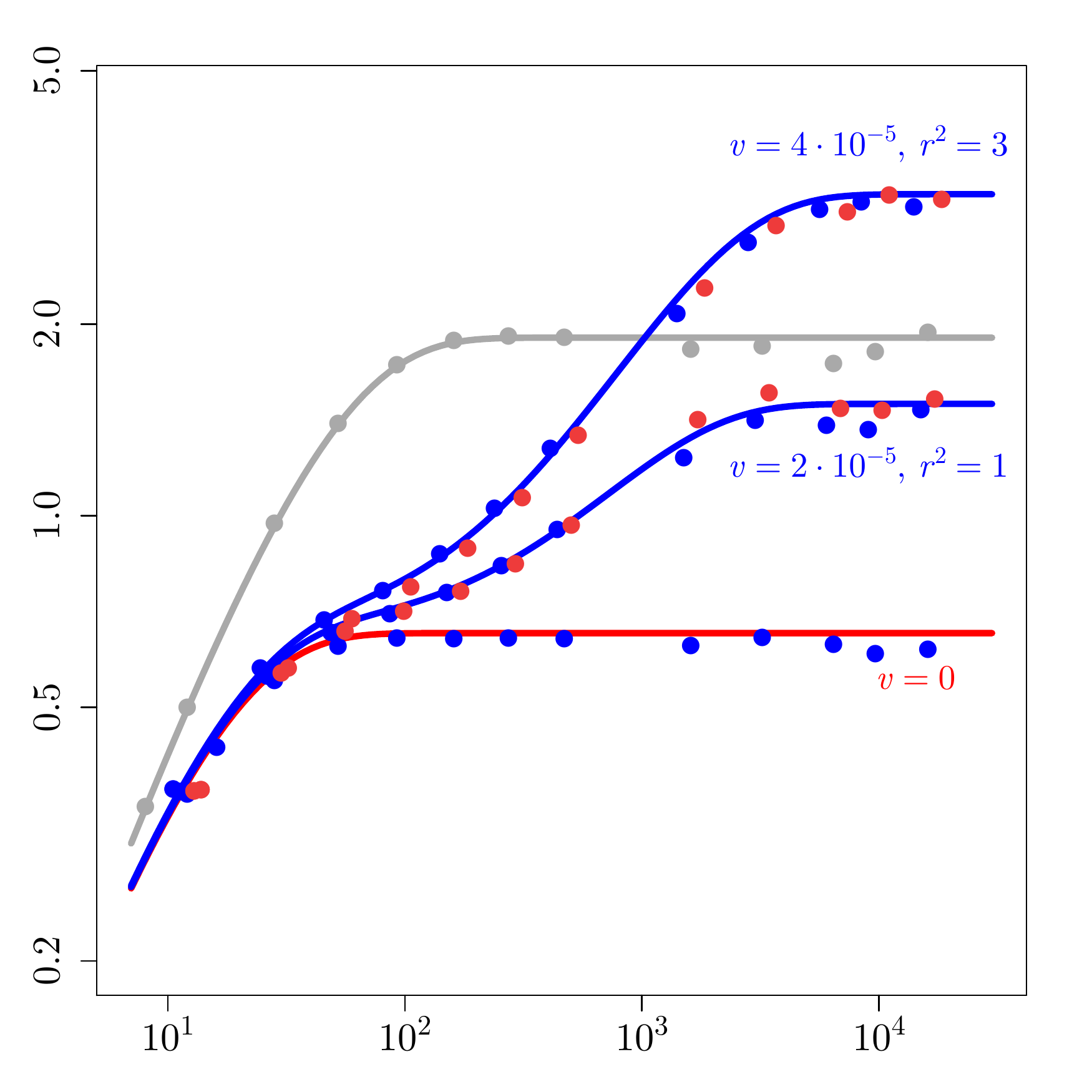} \\  
& \hspace{85pt} time, $\tau/N$ & &
\hspace{85pt} time, $\tau/N$ \\
&\fig{c} && \fig{d}  \\
\begin{sideways} \hspace{50pt} scaled genetic load, $\ql 2N L \qr$ \end{sideways} &
\includegraphics[width=.40\textwidth]{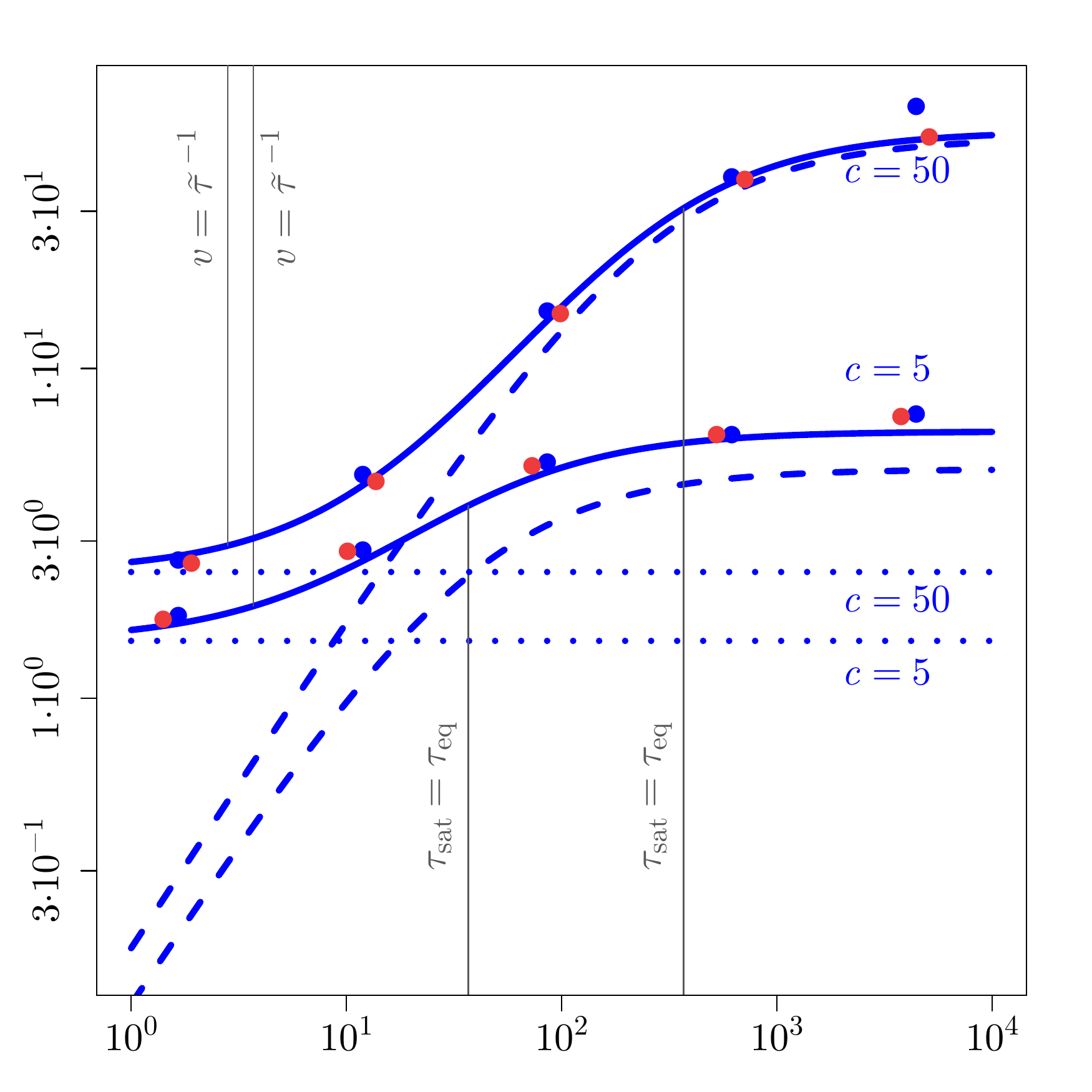} &
\begin{sideways} \hspace{50pt} scaled fitness flux, $\ql 2N \phi \qr/\mu$ \end{sideways}& 
\includegraphics[width=.40\textwidth]{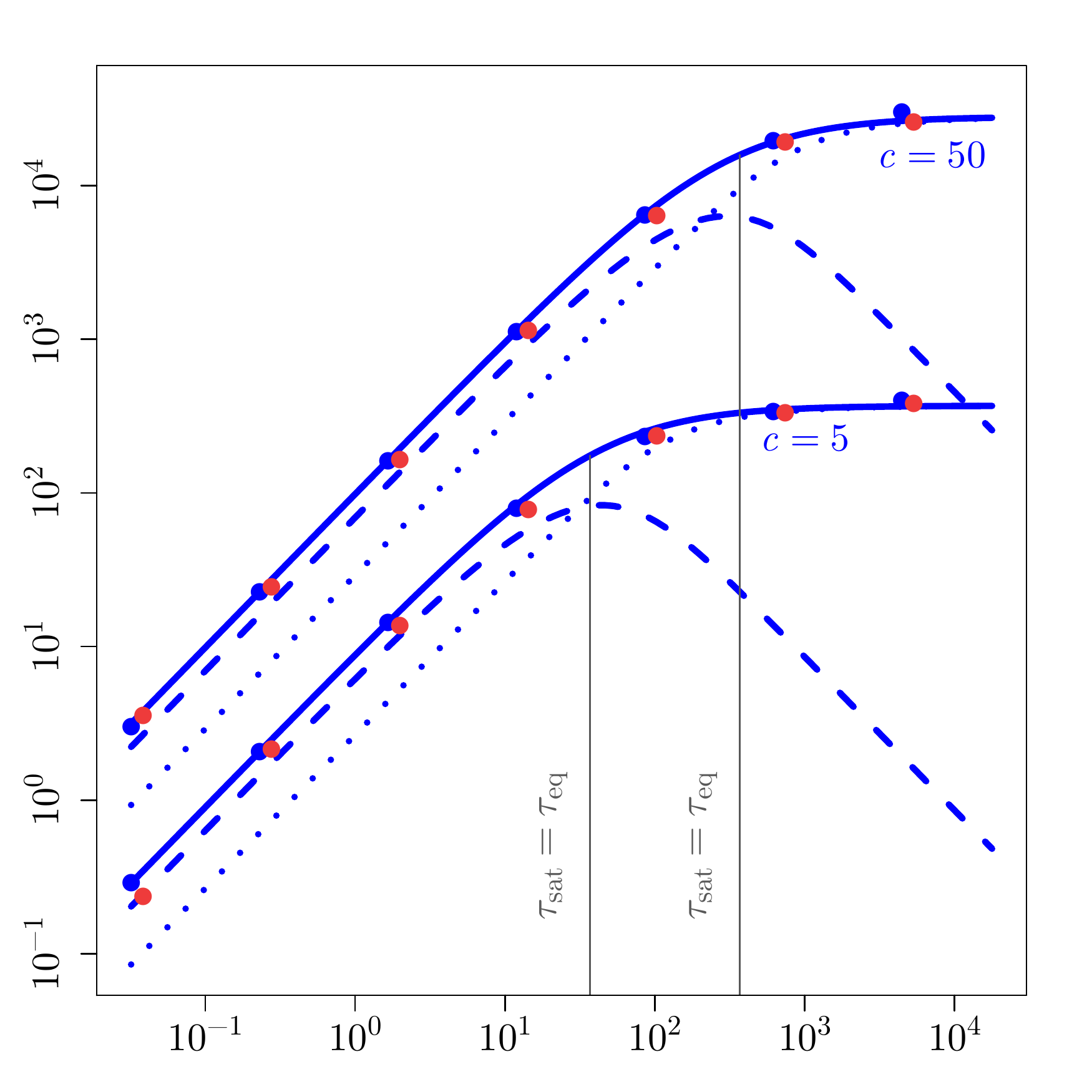}\\
&  \hspace{45pt} scaled driving rate, $\v /\mu$ & & 
\hspace{45pt} scaled driving rate, $\v /\mu$
\end{tabular}
\caption{{\bf Trait evolution under free recombination.} 
(a,b)~The scaled average divergence $\langle d^{(1)} \rangle(\tau)$ is shown as a function of the scaled divergence time $\tau/N$ for  three cases: neutral evolution ($c = 0$; grey lines), conservation in a static fitness landscape ($c = 1$, $\v = 0$; red line), and adaptation in a macro-evolutionary fitness seascape ($c = 1$, $\v > 0$; blue lines). The analytical results of eq.~(\ref{dkappa}) (lines) are compared to simulation results for evolution with free recombination in diffusive and punctuated fitness seascapes (blue and red dots, respectively). The analytical value of $\langle \delta \rangle$ is taken from eq.~(69) of ref.~\cite{Nourmohammad2012}; the other parameters are as in Fig.~\ref{fig:acfoverDlong}. 
(c)~Scaled genetic load $2NL$ (full lines), adaptive load $2NL_{\rm ad}$ (dashed lines),  and equilibrium load $2N L_\eq$ (dotted lines), plotted against the scaled driving rate $\v/\mu$. 
The other parameters are as in Fig. \ref{loadflux}(a). 
(d)~Scaled fitness flux $\langle 2N \phi \rangle$ and its components $\ql 2N\phi_{\rm micro}\qr$ and  $\ql 2N\phi_{\rm macro }\qr$ (with decomposition constant $k = 2$), plotted against the scaled driving rate $\v/\mu$. The other parameters are as in Fig. \ref{loadflux}(b). 
}
\label{fig:freerec}
\end{figure}

The evolutionary statistics of the trait mean depends weakly but systematically on the recombination rate; this dependence arises because the mean diversity $\langle \Delta \rangle$ enters the quasi-neutral dynamics of $\Gamma$~\cite{Nourmohammad2012}. To simulate evolution with a finite recombination rate $\rho$, we recombine  the genomes of pairs of individuals with probability $\epsilon \rho$ at a single random crossover position of the genome. For the simulation of free recombination, we randomly shuffle the alleles $a_{i}^{1}, . . . , a_{i}^{N}$ between the individuals at each genomic site $i$ and in each generation. Analytical and numerical results for the scaled divergence $\langle d^{(1)} \rangle (\tau)$, the  scaled genetic load $2NL$, and the scaled fitness flux $\langle 2N\phi \rangle / \mu$ under free recombination are shown in Fig.~\ref{fig:freerec}; these  should be compared with the corresponding results for non-recombining populations in Figs.~3, 4, and 5.   

\begin{figure}[t]
\centering
\begin{tabular}{rlll}
& \fig{a} & \fig{b} & \fig{c}\\ 
\begin{sideways} \hspace{10pt} 
 divergence/diversity ratio, $ \Omega(\tau)$ \end{sideways} &
 \includegraphics[width=0.3\textwidth]{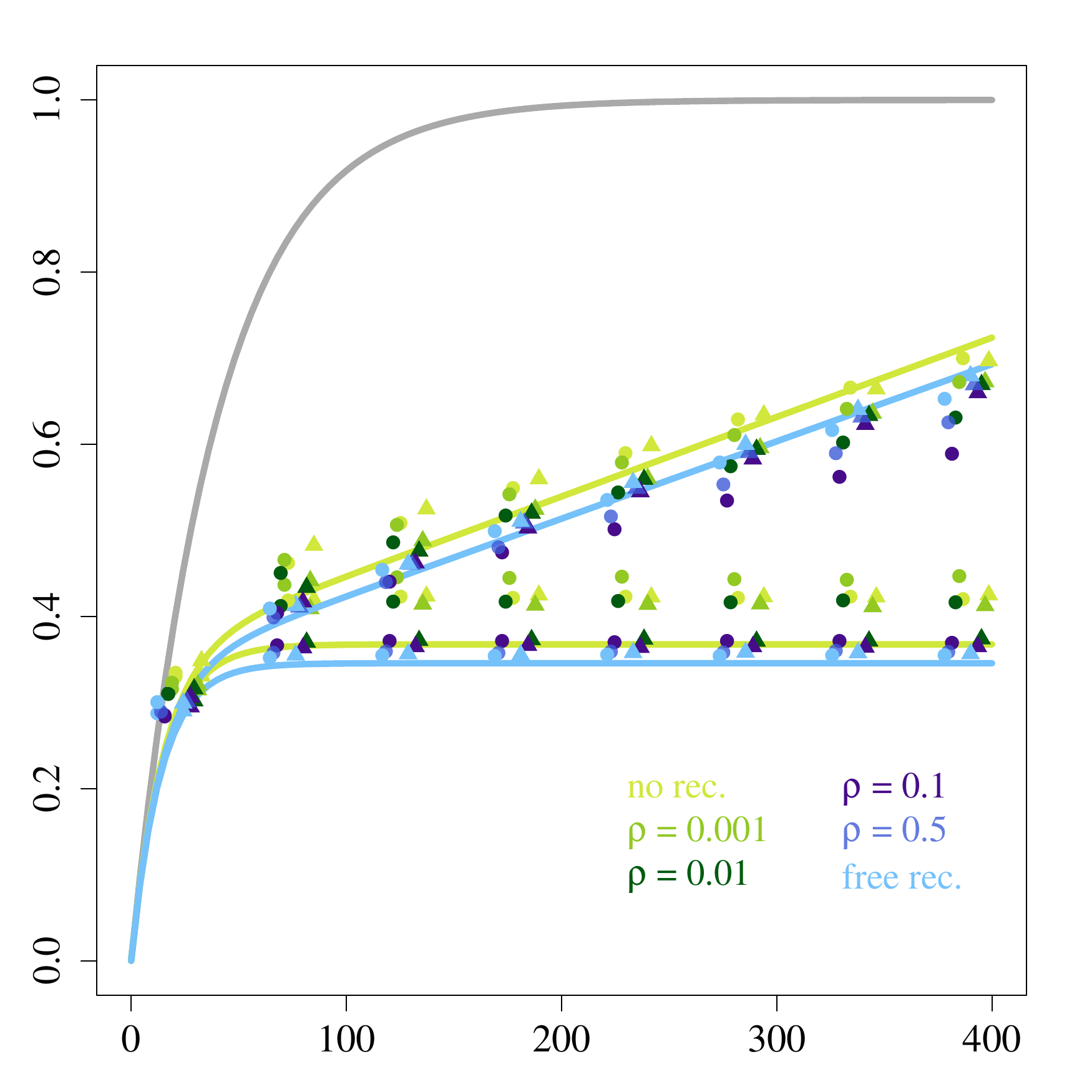}
 & 
\includegraphics[width=0.3\textwidth]{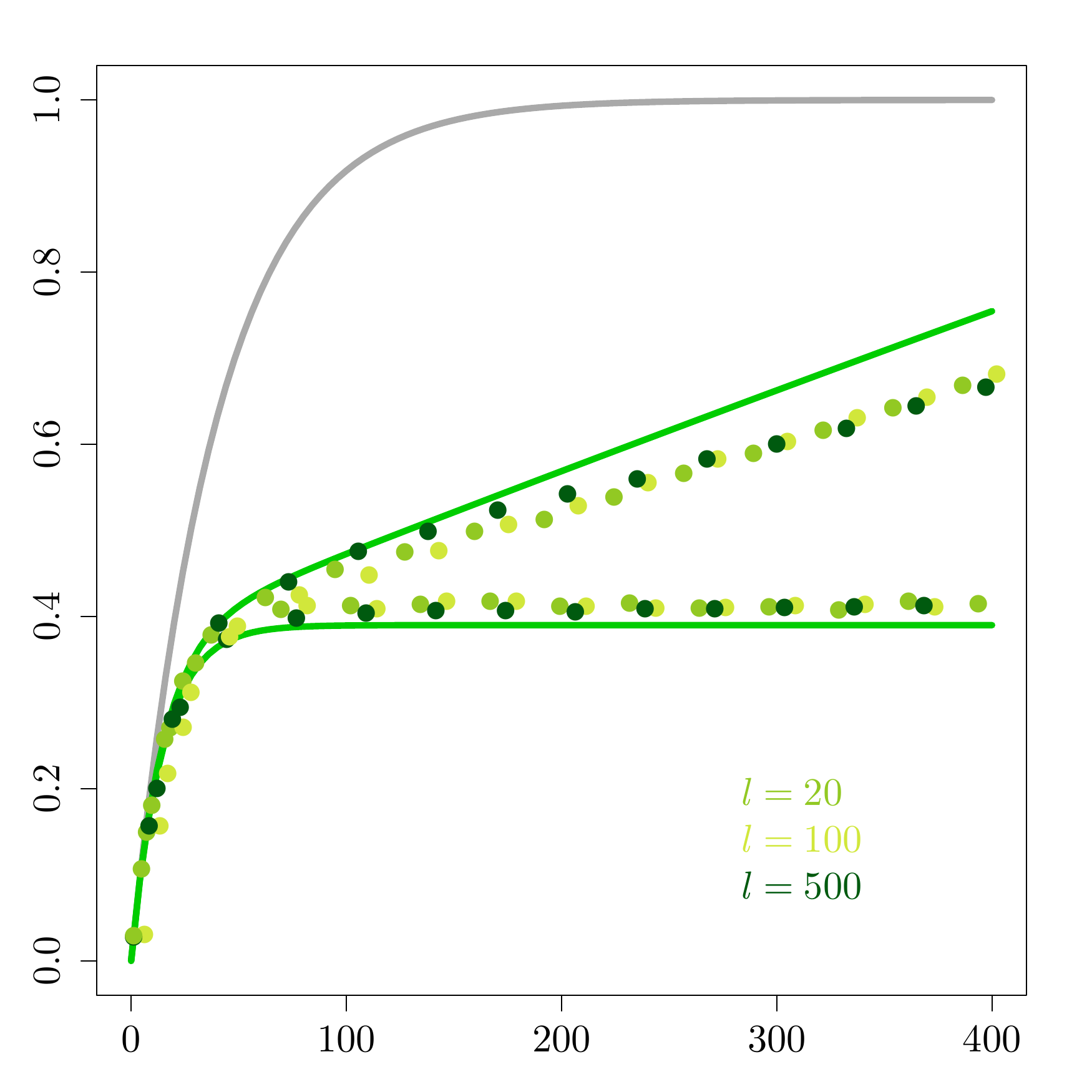}
& 
\includegraphics[width=0.3\textwidth]{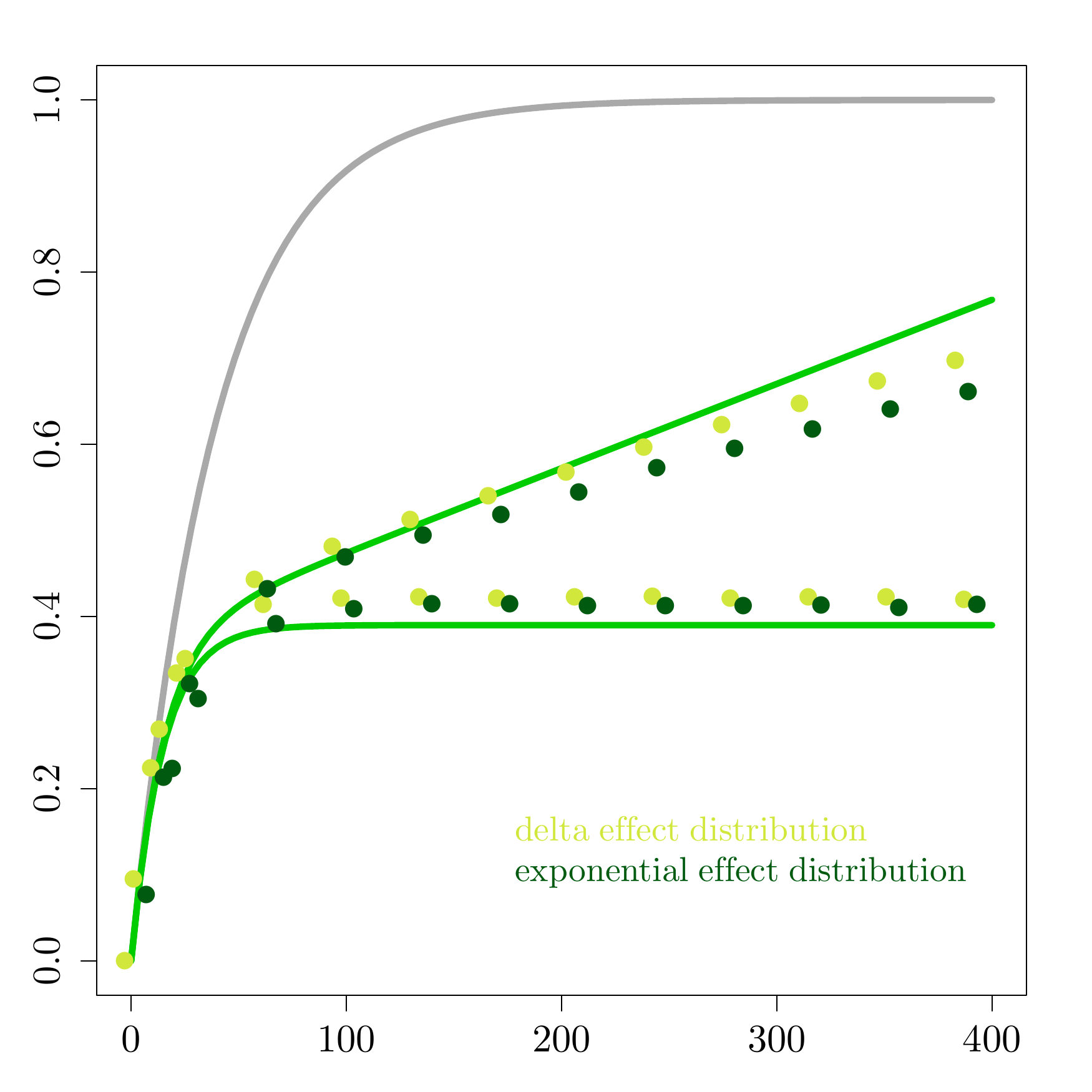}
\\ 
&\hspace{45pt}  time, $\tau/N$ &
\hspace{45pt} time, $\tau/N$ &
\hspace{45pt} time, $\tau/N$
\end{tabular} 
\caption{{\bf Universality of the divergence/diversity ratio $\Omega(\tau)$.} Numerical results for the evolution in fitness seascapes ($c =1$, $\v=4\cdot10^{-5}$,
upper lines and dots) and fitness landscapes ($c =1$, $\v = 0$, lower lines and dots) under different molecular conditions are compared to the analytical solutions for nonrecombining ($\rho=0$) and free-recombining ($\rho\rightarrow\infty$) genomes.
(a)~Evolution with different recombination rates (color-coded dots for diffusive seascapes and triangles for punctuated seascapes). 
(b)~Evolution with different numbers $\ell$ of constitutive sites in nonrecombining populations. 
(c)~Evolution wih different effect distributions. The trait amplitudes $E_i$ ($i =1, \dots, \ell$) are drawn from an exponential distribution with expectation value $1/\sqrt{2}$ and from a delta distribution (all sites have amplitude $E_i=1$).  }
\label{fig:rhos}
\end{figure}

Universality is the (approximate) independence of a summary trait observable from details of the trait's genomic encoding and of its molecular evolution~\cite{Nourmohammad2013}. In Fig.~\ref{fig:rhos}, we report three universality tests for the divergence-diversity ratio $\Omega^{(1)} (\tau)$. First, simulations show that the $\Omega$ statistics depends only weakly on the recombination rate throughout the crossover between asexual evolution ($\rho = 0$) and free recombination $(\rho \to \infty$). Second, the $\Omega$ ratio is invariant under variations in the number of constitutive genomic sites, $\ell$, at constant selection parameters $c$ and $\v$. Third, this ratio is also invariant under variations of the phenotypic effect sizes $E_i$ at these sites; this is tested by comparing simulations for two distributions of effect sizes. 

\end{appendix}

\end{document}